\documentclass[10pt,showpacs,amsmath,amssymb]{revtex4}

\usepackage{times}
\usepackage{graphicx}
\usepackage{bbold}
\usepackage{epsfig}
\usepackage{rotating}
\usepackage{graphpap}

\usepackage[]{graphicx}
\chardef\bslash=`\\ 

\begin{document}
\setlength{\unitlength}{1mm}

\pacs{05.45.-a, 87.18.Hf, 89.75.Fb\\[0.3cm]
{\it Dedicated to Hermann Haken on the occasion of his 80th birthday.}}

\title{Solutions of the H{\"a}ussler-von der Malsburg Equations in Manifolds with Constant Curvatures}

\author{M. G{\"u}{\ss}mann\footnote{E-mail: {\sf martin.guessmann@itp1.uni-stuttgart.de}}}
\affiliation{1.$\,$Institut f{\"u}r Theoretische Physik, Universit{\"a}t Stuttgart, Pfaffenwaldring 57, 70569 Stuttgart, Germany}
\author{A. Pelster\footnote{Corresponding author \quad E-mail: {\sf axel.pelster@uni-due.de}}}
\affiliation{Fachbereich Physik, Campus Duisburg, Universit{\"a}t Duisburg-Essen, Lotharstrasse 1, 47048 Duisburg, Germany}
\author{G. Wunner\footnote{E-mail: {\sf guenter.wunner@itp1.uni-stuttgart.de}}}
\affiliation{1.$\,\,$Institut f{\"u}r Theoretische Physik, Universit{\"a}t Stuttgart, Pfaffenwaldring 57, 70569 Stuttgart, Germany}

\begin{abstract}
We apply generic order parameter equations for the emergence of retinotopy
between manifolds of different geometry 
to one- and two-dimensional Euclidean and spherical manifolds. 
To this end we elaborate both a linear and a nonlinear synergetic analysis which results in 
order parameter equations for the dynamics of connection weights between two 
cell sheets. 
Our results for strings are analogous to those for discrete 
linear chains obtained previously by H{\"a}ussler and von der Malsburg. 
The case of planes turns out to be more involved as the two dimensions do not decouple in
a trivial way. However, superimposing two modes under suitable conditions 
provides a state with a pronounced retinotopic character. 
In the case of spherical manifolds we show that the order
parameter equations provide stable stationary solutions 
which correspond to retinotopic modes. 
A further analysis of higher modes furnishes proof
that our model describes the emergence of a perfect one-to-one retinotopy between two spheres. 
\end{abstract}
\maketitle
\section{Introduction}
In a preceding paper \cite{gpw1} we have analyzed
a general model for the formation of retinotopic projections
which is independent of geometry and dimension. In this paper we present
applications of the general model, viz Euclidean and spherical geometries in one and two
dimensions. To put these investigations into perspective we briefly recall their physiological motivations.
But let us stress at the outset that our primary objective is not the biological modelling
of retinotopic projections but the systematic analysis of a particular model
thereof from a nonlinear dynamics point of view.
In the course of ontogenesis of vertebrate animals well-ordered neural connections 
are established between retina and tectum, a part of the brain which plays an important 
role in processing optical information. At an initial stage of ontogenesis, the 
ganglion cells of the retina have random synaptic contacts with the tectum. In the 
adult animal, however, neighbouring retinal cells project onto neighbouring cells of the tectum \cite{goodhill}.
A detailed analytical treatment by H{\"a}ussler and von der Malsburg was able to describe 
the generation of such retinotopic states from an undifferentiated initial state as a 
self-organization process \cite{Malsburg}. In that work retina and tectum were treated 
as one-dimensional discrete cell arrays. The dynamics of the connection weights between 
retina and tectum was assumed to be governed by the H{\"a}ussler-von der Malsburg equations which 
are based on modelling the interplay between cooperative and competitive interactions 
of the individual synaptic contacts. The nonlinear analysis was performed using the methods 
of synergetics, which provides effective analytical methods to study self-organization 
processes in complex systems \cite{Haken1,Haken2}.\\

Obviously, the description of cell sheets as linear chains with the same number of 
cells is an inadequate approach to the real biological situation. 
In a preceding paper we generalized 
the underlying H{\"a}ussler-von der Malsburg equations to {\it continuous} manifolds of {\it arbitrary} geometry 
and dimension \cite{gpw1}. 
We performed an extensive synergetic analysis of these generalized H{\"a}ussler-von der Malsburg equations. 
The resulting generic order 
parameter equations represented a central new result, and can now
serve as a 
starting point to analyze in detail the self-organized emergence of 
one-to-one mappings in cell 
arrays of different geometries. A short review of our generalization of the H{\"a}ussler-von der Malsburg 
equations and the results of the corresponding synergetic analysis is provided in Sec.~\ref{generalmodel}. In the subsequent two Sections we focus on one- and two-dimensional 
Euclidean manifolds. We show in Sec.~\ref{strings} that the treatment of 
strings yields results which are analogous to those obtained for discrete linear chains in 
Ref.~\cite{Malsburg}, i.e.~our model includes the special case discussed by H{\"a}ussler and 
von der Malsburg. However, our synergetic analysis is more general. Instead of discrete cell arrays 
with the same number of cells, we consider continuously distributed cells on strings 
of different lengths \cite{tueb}. Furthermore, 
we do not restrict our investigations to monotonically 
decreasing cooperativity functions of strings.
We investigate under what circumstances non-retinotopic modes become unstable and
destroy retinotopic order. 
We show in Sec.~\ref{planes} that our 
generic order parameter equations also provide a suitable 
framework to describe the emergence of retinotopy between planes.
For a certain superposition of two modes we demonstrate that taking into account the contribution of the higher modes
leads to a sharpening of the retinotopic character of the projection between the two cell sheets.
Finally, we analyse in Sec.~\ref{spheres} the formation of retinotopic projections
for the biologically relevant situation of spherical geometries, i.e. manifolds with {\it positive constant curvature}. It turns out that the case of spheres
exhibits remarkable similarities with the analysis of strings, especially
regarding the generation of 1-1-retinotopic projections. 
\begin{figure}[t!]
\centerline{\includegraphics[scale=0.6]{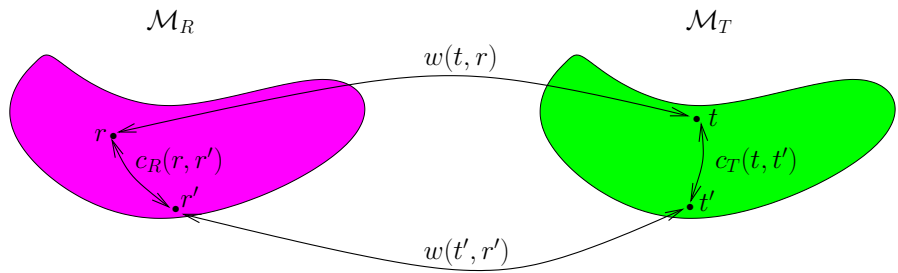}}
\caption{\label{manifolds} \small Retina and tectum are represented as manifolds 
${\cal{M}}_R$ and ${\cal{M}}_T$, respectively, which are connected by positive connection 
weights $w(t,r)$. The connectivity within each manifold is represented by cooperativity
functions $c_R(r,r')$ and $c_T(t,t')$.}
\end{figure}
\section{The General Model} \label{generalmodel}

To make the present paper self-contained, we briefly review 
the essential results of our general model for the self-organized emergence of retinotopic 
projections between manifolds of different geometry in Ref.~\cite{gpw1}.
The two cell sheets, retina and tectum, are represented by general manifolds ${\cal M}_R$ 
and ${\cal M}_T$, respectively. Every ordered pair $(t,r)$ with 
$t\in {\cal M}_T,\,r\in {\cal M}_R$ is connected by a connection weight $w(t,r)$ 
as is illustrated in Fig.~\ref{manifolds}. The equations of evolution of these connection
weights are assumed to be given by a generalization of the {\it H{\"a}ussler-von der Malsburg equations} 
\begin{equation} 
\label{HAUS}
\dot w(t,r)=f(t,r,w)-\frac{w(t,r)}{2 M_T} 
\hspace*{2mm} \int \! dt'\,f(t',r,w)
-\frac{w(t,r)}{2 M_R}\hspace*{2mm} 
\int \! dr'\,f(t,r',w)\,,
\end{equation}
where the first term on the right-hand side describes cooperative synaptic growth processes and the other terms stand for corresponding competitive growth processes. 
Here $M_T\,,\,M_R$ denote the magnitudes of the manifolds, the total growth rates are defined by
\begin{equation}
\label{GRO}
f(t,r,w)=\alpha+w(t,r) \int \! dt'
\int \! dr' c_T(t,t')\,
c_R(r,r')\,w(t',r')\,, 
\end{equation}
and $\alpha$ is the global growth rate of new synapses onto the tectum which 
represents the control parameter of our system. The cooperativity functions $c_T(t,t')$, 
$c_R(r,r')$ represent the neural connectivity within each manifold. We assume that 
they are positive, symmetric with respect to their arguments and normalized. 
The cooperation strength depends on the distance between two points of the manifold. 
This requires a measure of distance, i.e.~metrics, which in turn define Laplace-Beltrami 
operators on the manifolds. Their eigenvalue problems yield a complete orthonormal system 
$\psi_{\lambda_T}(t)\,,\psi_{\lambda_R}(r)$, and the generalized H{\"a}ussler-von der Malsburg equations are most conveniently 
transformed to this new basis. For example, the cooperativity functions are expanded 
in terms of these functions as follows:
\begin{eqnarray}
\label{CEX}
c_T(t,t')=\sum_{\lambda_T}
f_{\lambda_T} \psi_{\lambda_T} (t)
\psi_{\lambda_T}^{*}(t')\,,\hspace*{1cm}
c_R(r,r')=\sum_{\lambda_R}
f_{\lambda_R} \psi_{\lambda_R} (r)
\psi_{\lambda_R}^{*}(r')\,.
\end{eqnarray}
The initial state of ontogenesis with randomly distributed synaptic contacts is 
described by the stationary uniform solution of the generalized H{\"a}ussler-von der Malsburg equations 
$w_0(t,r)=1$. Its stability is analyzed by linearizing the H{\"a}ussler-von der Malsburg equations (\ref{HAUS})
with respect to the deviation $v(t,r)=w(t,r)-w_0(t,r)$.
The resulting linearized equations read
$\dot v(t,r)=\hat{L}(t,r,v)$
with the linear operator 
\begin{eqnarray}
\label{Loperator}
\hat{L}(t,r,v)&=&-\alpha v (t,r)+\int\! dt'\int\! dr'\, c_T(t,t') \, c_R(r,r')\, v(t',r')
\nonumber\\
&&-\frac{1}{2 M_T}\int\! dt' \left[v(t',r) + 
\int\! dt''\int\! dr''\, c_T(t',t'') \, c_R(r,r'')\, v(t'',r'')
\right]
\nonumber \\
&&-\frac{1}{2 M_R}\int\! dr' \left[v(t,r') + 
\int\! dt''\int\! dr''\, c_T(t,t'') \, c_R(r',r'')\, v(t'',r'')
\right]
\,.
\end{eqnarray}
The eigenvalue problem of 
the linear operator (\ref{Loperator}) is solved by the eigenfunctions
\begin{equation} \label{eigenf}
v_{\lambda_T \lambda_R}(t,r)=\psi_{\lambda_T}(t)\psi_{\lambda_R}(r)
\end{equation}
and the following spectrum of eigenvalues:
\begin{equation} \label{ewsaite}
\Lambda_{\lambda_T \lambda_R}=\left\{\begin{array}{cc}
-\alpha-1&\quad \lambda_T=\lambda_R=0\vspace{2mm}\\
-\alpha+\frac{1}{2}(f_{\lambda_T}^T f_{\lambda_R}^R-1)&\quad \lambda_T=0,\lambda_R\not=0;\lambda_R=0,\lambda_T\not=0\vspace{2mm}\\
-\alpha+f_{\lambda_T}^T f_{\lambda_R}^R&\quad  \mbox{ otherwise. }
\end{array} \right.
\end{equation}
The eigenvalue with the largest real part is given by $\Lambda_{\rm max}=-\alpha+f_{\lambda_T^u}^T f_{\lambda_R^u}^R$,
where $\lambda_T^u$, $\lambda_R^u$ denote all those eigenvalues which could become unstable simultaneously.
Thus, the instability takes place when the global growth rate reaches
its critical value $\alpha_c=\mbox{Re}\,(f_{\lambda_T^u}^T f_{\lambda_R^u}^R)$.\\

The linear stability analysis motivates to treat the nonlinear H{\"a}ussler-von der Malsburg equations (\ref{HAUS})
near the instability by decomposing the deviation $v(t,r)=w(t,r)-w_0(t,r)$
in unstable and stable contributions according to
$v(t,r) = U(t,r)+S(t,r)$.
With Einstein's sum convention we have for the unstable modes
\begin{equation}
\label{UEXP}
U(t,r)=
U_{\lambda_T^u \lambda_R^u} \psi_{\lambda_T^u} (t) \psi_{\lambda_R^u} (r) \, ,
\end{equation}
and, correspondingly, 
\begin{equation}
\label{SEXP}
S(t,r)=
S_{\lambda_T \lambda_R}\, \psi_{\lambda_T} (t) \psi_{\lambda_R} (r)
\end{equation}
represents the contribution of the stable modes. 
Note that the summation in (\ref{SEXP}) is performed over all parameters $(\lambda_T ;\lambda_R )$ 
except for $(\lambda_T^u ;\lambda_R^u )$, i.e. from now on the parameters $(\lambda_T ;\lambda_R)$ stand for the stable modes alone. 
With the help of the 
slaving principle of synergetics the original high-dimensional system can be reduced 
to a low-dimensional one which only contains the unstable amplitudes. The general form 
of the resulting order parameter equations is independent of the geometry of the problem 
and reads
\begin{eqnarray}
\dot U_{\lambda_T^u \lambda_R^u} &=&\Lambda_{\lambda_T^u \lambda_R^u} \,
U_{\lambda_T^u \lambda_R^u} +
A^{\lambda_T^u, \lambda_T^u{}' \lambda_T^u{}''}_{\lambda_R^u, \lambda_R^u{}' \lambda_R^u{}''}
\,U_{\lambda_T^u{}' \lambda_R^u{}'} \, U_{\lambda_T^u{}''\lambda_R^u{}''}\nonumber\\
 & &+ B^{\lambda_T^u, \lambda_T^u{}' \lambda_T^u{}'' \lambda_T^u{}'''}_{\lambda_R^u{},
\lambda_R^u{}' \lambda_R^u{}'' \lambda_R^u{}'''} U_{\lambda_T^u{}'\lambda_R^u{}'}
\, U_{\lambda_T^u{}'' \lambda_R^u{}''}\, U_{\lambda_T^u{}''' \lambda_R^u{}'''} \,.
\label{OPE}
\end{eqnarray}
It contains, as is typical, a linear, a quadratic, and a cubic term of the order parameters. 
The corresponding coefficients can be expressed in terms
of the expansion coefficients $f_{\lambda_T}$, $f_{\lambda_R}$
of the cooperativity functions (\ref{CEX}) and integrals over products of the eigenfunctions
$\psi_{\lambda_T}(t)$, $\psi_{\lambda_R}(r)$:
\begin{eqnarray} 
I^{\lambda}_{\lambda^{(1)} \lambda^{(2)} \ldots  \lambda^{(n)}}
&=&\hspace*{1mm}\int\! dx\,
\psi_{\lambda}^{*}(x)\,\psi_{\lambda^{(1)}}(x)\,\psi_{\lambda^{(2)}}(x)\,\cdots\,
\psi_{\lambda^{(n)}}(x)\,,\label{abkurzI}\\
J_{\lambda^{(1)} \lambda^{(2)} \ldots  \lambda^{(n)}}
&=&\hspace*{1mm}\int\!dx\,
\psi_{\lambda^{(1)}} (x)\, \psi_{\lambda^{(2)}} (x)\,\cdots\,
\psi_{\lambda^{(n)}} (x)\,. \label{abkurzJ}
\end{eqnarray}
The quadratic coefficients read
\begin{eqnarray}
\label{AA}
A^{\lambda_T^u, \lambda_T^u{}' \lambda_T^u{}''}_{\lambda_R^u, \lambda_R^u{}' \lambda_R^u{}''}  = f_{\lambda_T^u{}''} \,
f_{\lambda_R^u{}''} \, I^{\lambda_T^u}_{\lambda_T^u{}' \lambda_T^u{}''}
\, I^{\lambda_R^u}_{\lambda_R^u{}' \lambda_R^u{}''} \,,
\end{eqnarray}
whereas the cubic coefficients are
\begin{eqnarray}
&&\hspace{-1.2cm}B^{\lambda_T^u, \lambda_T^u{}' \lambda_T^u{}'' \lambda_T^u{}'''}_{\lambda_R^u{}, \lambda_R^u{}' \lambda_R^u{}'' \lambda_R^u{}'''}
= - \frac{1}{2}\,f_{\lambda_T^u{}'''} \, f_{\lambda_R^u{}'''}
\left( \frac{1}{M_R}\,I^{\lambda_T^u}_{\lambda_T^u{}' \lambda_T^u{}'' \lambda_T^u{}'''}
\,\delta_{\lambda_R^u \lambda_R^u{}'} \,J_{\lambda_R^u{}'' \lambda_R^u{}'''}
+ \frac{1}{M_T}\,I^{\lambda_R^u}_{\lambda_R^u{}' \lambda_R^u{}'' \lambda_R^u{}'''} \delta_{\lambda_T^u \lambda_T^u{}'}
\right.\nonumber\\
&&\hspace{-1.05cm}\times J_{\lambda_T^u{}'' \lambda_T^u{}'''} \bigg)
+\left\{ \left[ f_{\lambda_T} \, f_{\lambda_R} +f_{\lambda_T^u{}'} \, f_{\lambda_R^u{}'} \, \right]
I^{\lambda_T^u}_{\lambda_T^u{}' \lambda_T} \,
I^{\lambda_R^u}_{\lambda_R^u{}' \lambda_R}
- \frac{1}{2} \,
\left[ \,\frac{1}{\sqrt{M_T}}\,\delta_{\lambda_T 0}\, \delta_{\lambda_T^u \lambda_T^u{}'}  \,
\left( 1+ f_{\lambda_R} \right)
I^{\lambda_R^u}_{\lambda_R^u{}' \lambda_R}
\right.\right. \nonumber \\ && \left. \left.
\hspace{-1.05cm}+\,\frac{1}{\sqrt{M_R}}\,\delta_{\lambda_R 0}\,\delta_{\lambda_R^u \lambda_R^u{}'}
\, \left( 1+f_{\lambda_T}\right)
\, I^{\lambda_T^u}_{\lambda_T^u{}' \lambda_T}
\right] \right\} H^{\lambda_T \lambda_R}_{\lambda_T^u{}'' \lambda_R^u{}'',\lambda_T^u{}''' \lambda_R^u{}'''} \,
\,.
\label{BB}
\end{eqnarray}
As is common in synergetics, the cubic coefficients (\ref{BB}) consist in general 
of two parts, one stemming from the order parameters themselves and the other 
representing the influence of the center manifold $H$ on the order parameter dynamics according to
\begin{eqnarray}
\label{HN}
S_{\lambda_T \lambda_R} = 
H^{\lambda_T \lambda_R}_{\lambda_T^u \lambda_R^u,\lambda_T^u{}' \lambda_R^u{}'} \,
U_{\lambda_T^u \lambda_R^u} U_{\lambda_T^u{}' \lambda_R^u{}'} \, .
\end{eqnarray}
Here the center manifold coefficients $H^{\lambda_T \lambda_R}_{\lambda_T^u \lambda_R^u,\lambda_T^u{}' \lambda_R^u{}'}$ are 
defined by 
\begin{eqnarray}
\label{HNN}
H^{\lambda_T \lambda_R}_{\lambda_T^u \lambda_R^u,\lambda_T^u{}' \lambda_R^u{}'}  &=& 
\frac{f_{\lambda_T^u{}'} \, f_{\lambda_R^u{}'}}{\Lambda_{\lambda_T^u \lambda_R^u} +\Lambda_{\lambda_T^u{}' \lambda_R^u{}'}- \Lambda_{\lambda_T \lambda_R} }
\Bigg[ I^{\lambda_T}_{\lambda_T^u \lambda_T^u{}'} \, 
I^{\lambda_R}_{\lambda_R^u \lambda_R^u{}'} 
\nonumber \\&& 
- \frac{1}{2} \, \left( \frac{1}{\sqrt{M_T}}\,J_{\lambda_T^u \lambda_T^u{}'} 
\,I^{\lambda_R}_{\lambda_R^u \lambda_R^u{}'} \,\delta_{\lambda_T 0}  
+ \frac{1}{\sqrt{M_R}}\, J_{\lambda_R^u \lambda_R^u{}'} \,I^{\lambda_T}_{\lambda_T^u \lambda_T^u{}'} 
\,\delta_{\lambda_R 0} \right) \Bigg] \,. 
\label{QRES}
\end{eqnarray}
The order parameter equations (\ref{OPE}) for the generalized H{\"a}ussler-von der Malsburg equations (\ref{HAUS})
can now serve as a starting point
for analysing the self-organized formation of retinotopic projections between manifolds
of different geometry. 
\section{Strings} \label{strings}
In this section we specialize the generic order parameter equations (\ref{OPE}) to 
one-dimensional Euclidean manifolds of strings with different lengths $L_T$ and $L_R$. 
We start with introducing the eigenfunctions in Subsec.~\ref{stringef}.
In Subsec.~\ref{stringsa} we observe
that the quadratic term vanishes and derive selection rules for the appearance of cubic
terms. In this way we essentially simplify the calculation of order parameter equations as
compared with Ref.~\cite{Malsburg}. Furthermore, we show that the order parameter equations
represent a potential dynamics, and determine the underlying potential in Subsec.~\ref{stringcop}. A subsequent
transformation from complex to real order parameters in Subsec.~\ref{stringrop} leads to constant phase-shift angles.
Thus, in Subsec.~\ref{konstantephasenwinkel} we reduce the order parameter dynamics to two variables which correspond to the
amplitudes of two diagonal modes. These two modes compete with each other, until one of them
vanishes. Within the potential picture this means that the stable uniform
state becomes unstable and the system settles in one of the two potential minima, as is discussed in Subsec.~\ref{saitenpoteigen}.
After one of the diagonal modes has won, only such modes are excited which contribute
to the sharpening of the diagonal. Approximately solving the H{\"a}ussler-von der Malsburg equations in Subsec.~\ref{stringotor} leads to
the following scenario: Above a critical global growth rate $\alpha_c$ the uniform state
$w_0(t,r)=1$ is stable. By decreasing the control parameter $\alpha$, the projection gets
sharper and sharper. Finally, if there is no global growth rate of new synapses any more,
i.e.~$\alpha = 0$, the connection weights are given by Dirac's delta function. Thus, a perfect
one-to-one retinotopic state is realized. We conclude this discussion of strings
with comparing our results with the corresponding analysis of discrete linear chains in Subsec.~\ref{stringclc}.
\subsection{Eigenfunctions} \label{stringef}
The magnitudes of the manifolds 
${\cal M}_T$ and ${\cal M}_R$ are given by $M_T=L_T$ and $M_R=L_R$, 
respectively. To avoid problems at the boundaries, we assume periodic boundary conditions, 
i.e.~we consider retina and tectum to be rings with circumferences $L_T$ and $L_R$, respectively.
The eigenvalue problem of the Laplace-Beltrami operator for both manifolds reads
\begin{equation}
\frac{\partial^2}{\partial x^2}\,\psi_{\lambda}(x)=\chi_{\lambda}\psi_{\lambda}(x)\,,
\end{equation}
with $x=t,r$, respectively. Using the boundary condition $\psi_{\lambda}(x)=\psi_{\lambda}(x+L)$, 
this is solved by the eigenfunctions
\begin{equation} 
\label{saitenef}
\psi_{\lambda}(x)=\frac{1}{\sqrt L} \exp\left(i\frac{2\pi}{L}\lambda x\right)\,,
\end{equation}
where the eigenvalues are given by
$\chi_{\lambda}=-4\pi^2 \lambda^2/L^2$,
with $x \in [0,L)$ and $\lambda=0,\pm 1,\pm 2,\ldots\,$. Every eigenvalue $\chi_{\lambda}$, apart from 
the special case $\chi_0=0$, is two-fold degenerate. 
The eigenfunctions form a complete orthonormal system:
\begin{equation} \label{ortho}
\int\limits_0^{L}\!\! dx \, \psi_{\lambda}(x)\psi_{\lambda '}^*(x)=\delta_{\lambda \lambda '}\,,\quad \sum_{\lambda=-\infty}^{\infty} \psi_{\lambda}(x)\psi_{\lambda}^*(x ')=\delta(x-x')\,.
\end{equation}
Note that the orthonormality relation in (\ref{ortho}) follows directly by inserting 
(\ref{saitenef}), whereas the completeness relation  
is proven by taking into account the Poisson formula \cite{klein}.
The cooperativity functions only depend on the distance, which is given by the Euclidean 
distance $|x-x '|$, i.e.~$c(x,x')=c(x-x')$.
Their expansion in terms of the eigenfunctions (\ref{saitenef}) corresponds to the Fourier series 
\begin{equation} 
\label{1dct}
c(x-x')=\frac{1}{L}\sum_{\lambda=-\infty}^{\infty} f_{\lambda} \exp\left[i\frac{2\pi}{L}\lambda \left(x-x'\right)\right]
\,.
\end{equation}
The expansion coefficients $f_{\lambda}$ are independent of the sign of the parameters $\lambda$, 
i.e.~$f_{\lambda}=f_{-\lambda}$, as the cooperativity functions  are symmetric with 
respect to their arguments: $c(x-x')=c(x'-x)$.
\subsection{Synergetic Analysis} \label{stringsa}
To specialize the order parameter equations (\ref{OPE}) to the case of strings, we have 
to determine the integrals (\ref{abkurzI}) and (\ref{abkurzJ}) of products of eigenfunctions. 
With (\ref{saitenef}) we obtain 
\begin{eqnarray}
I_{\lambda^{(1)}\lambda^{(2)}\ldots\,\lambda^{(n)}}^{\lambda}&=&\left(\frac{1}{\sqrt{L}}\right)^{n-1}\delta_{\lambda^{(1)}+\lambda^{(2)}
+\ldots +\lambda^{(n)},\lambda}\,,\\
J_{\lambda^{(1)} \lambda^{(2)} \ldots\,\lambda^{(n)}}&=&\left(\frac{1}{\sqrt{L}}\right)^{n-2}\delta_{\lambda^{(1)}+\lambda^{(2)}
+\ldots+\lambda^{(n)},0}\,.\label{saitenj}
\end{eqnarray}
From these results one can immediately read off the special cases
\begin{equation} \label{1dgesamt}
I_{\lambda' \, \lambda''}^{\lambda}=\frac{1}{\sqrt{L}}\,\delta_{\lambda'+\lambda'',\lambda}\,,\quad 
I_{\lambda' \, \lambda'' \, \lambda'''}^{\lambda}=\frac{1}{L}\,\delta_{\lambda'+\lambda''+\lambda''',\lambda}\,,\quad
J_{\lambda' \, \lambda''}=\delta_{\lambda',-\lambda''}\,.
\end{equation}
In (\ref{AA}) the first integral of (\ref{1dgesamt}) occurs only with unstable values of $\lambda$. As they 
can only differ by their sign, it follows $\lambda'+\lambda''\not= \lambda$. Thus, we have $I_{\lambda' \, \lambda''}^{\lambda}=0$, 
i.e.~the quadratic term (\ref{AA}) in (\ref{OPE}) vanishes:
$A^{\lambda_T^u, \lambda_T^u{}' \lambda_T^u{}''}_{\lambda_R^u, \lambda_R^u{}' \lambda_R^u{}''}=0$.
To arrive at a more concise representation, we split the cubic contribution 
in (\ref{OPE}) in two terms according to
\begin{equation}
B^{\lambda_T^u, \lambda_T^u{}' \lambda_T^u{}'' \lambda_T^u{}'''}_{\lambda_R^u{},
\lambda_R^u{}' \lambda_R^u{}'' \lambda_R^u{}'''} U_{\lambda_T^u{}'\lambda_R^u{}'}
\, U_{\lambda_T^u{}'' \lambda_R^u{}''}\, U_{\lambda_T^u{}''' \lambda_R^u{}'''}=K_{1,\lambda_T^u \lambda_R^u}+K_{2,\lambda_T^u \lambda_R^u} \,.
\end{equation}
The first term $K_1$ takes into account the contribution of the order parameters themselves, 
while the second term $K_2$ represents the influence of the center manifold on the order 
parameter dynamics. Applying the integrals (\ref{1dgesamt}) leads to selection
rules for the appearance of cubic terms. It turns out that only those sums lead 
to non-vanishing contributions where
the sum of three unstable modes $\lambda^{u'}+\lambda^{u''}+\lambda^{u'''}$ coincides with another unstable
mode $\lambda^{u'}$. Thus, both for retina and tectum only the following combinations are
allowed: $(\lambda^{u'},\lambda^{u''},\lambda^{u'''}) = ( \lambda^u, \lambda^u,- \lambda^u)\, , ( \lambda^u, - \lambda^u,\lambda^u) \, , (- \lambda^u, \lambda^u,\lambda^u) \,$.
With this selection rule the first cubic term is given by
\begin{eqnarray}
K_{1,\lambda_T^u \lambda_R^u}&=&-\frac{f_{\lambda_T^u}^T f_{\lambda_R^u}^R}{2 L_T L_R}
\left[U_{\lambda_T^u \lambda_R^{u'}}U_{\lambda_T^{u''} \lambda_R^{u''}}U_{-\lambda_T^{u''} \lambda_R^{u'''}}
\delta_{\lambda_R^{u'}+\lambda_R^{u''}+\lambda_R^{u'''},\lambda_R^u}\right.\nonumber\\
& &\left.+U_{\lambda_T^{u'} \lambda_R^u}U_{\lambda_T^{u''} \lambda_R^{u''}}U_{\lambda_T^{u'''} -\lambda_R^{u''}}
\delta_{\lambda_T^{u'}+\lambda_T^{u''}+\lambda_T^{u'''},\lambda_T^u}\right]
\label{1dk1u}
\end{eqnarray}
and the second cubic term reads
\begin{eqnarray} \label{1d2ku}
&&\hspace{-1.3cm}K_{2,\lambda_T^u \lambda_R^u}=\frac{H_{\lambda_T' \lambda_R',\lambda_T^{u''} \lambda_R^{u''} \lambda_T^{u'''} \lambda_R^{u'''}}}{\sqrt{L_T L_R}}\,
U_{\lambda_T^{u'} \lambda_R^{u'}}U_{\lambda_T^{u''} \lambda_R^{u''}}U_{\lambda_T^{u'''} \lambda_R^{u'''}}\left\{(f_{\lambda_T'}^T 
f_{\lambda_R'}^R+f_{\lambda_T^u}^T f_{\lambda_R^u}^R)\delta_{\lambda_T^{u'}+\lambda_T',\lambda_T^u}\right.\nonumber\\
&&\hspace{-1.1cm}\times\delta_{\lambda_R^{u'}+\lambda_R',\lambda_R^u}
\left. -\frac{1}{2}
\left[(1+f_{\lambda_R'}^R)\delta_{\lambda_R^{u'}+\lambda_R',\lambda_R^u}\delta_{\lambda_T' 0}\delta_{\lambda_T^{u'}\lambda_T^u}
+(1+f_{\lambda_T'}^T)\delta_{\lambda_T^{u'}+\lambda_T',\lambda_T^u}\delta_{\lambda_R' 0}\delta_{\lambda_R^{u'}\lambda_R^u}\right]\right\}\,.
\end{eqnarray}
The latter depends on the center manifold, which follows from (\ref{HNN}), 
and (\ref{1dgesamt}) to be
\begin{eqnarray}
\hspace{-0.5cm}H_{\lambda_T \lambda_R,\lambda_T^{u''}\lambda_R^{u''}\lambda_T^{u'''}\lambda_R^{u'''}}&=&
\frac{f_{\lambda_T^u}^T f_{\lambda_R^u}^R}{\sqrt{L_T L_R} \left(2\Lambda_{\lambda_T^u \lambda_R^u}-\Lambda_{\lambda_T \lambda_R}\right)}\,
\left[\delta_{\lambda_T^{u''}+\lambda_T^{u'''},\lambda_T}\delta_{\lambda_R^{u''}+\lambda_R^{u'''},\lambda_R} \right. \nonumber \\
&&\hspace{-1.5cm}\left. -\frac{1}{2}\left(\delta_{\lambda_T^{u''},-\lambda_T^{u'''}}\delta_{\lambda_R^{u''}+\lambda_R^{u'''},\lambda_R}\delta_{\lambda_T 0}
+\delta_{\lambda_R^{u''},-\lambda_R^{u'''}}\delta_{\lambda_T^{u''}+\lambda_T^{u'''},\lambda_T}\delta_{\lambda_R 0}\right)\right]\,.
\label{1dzentrmann}
\end{eqnarray}
\subsection{Complex Order Parameters} \label{stringcop}

We can therefore conclude that the order parameter equations for strings have the form
\begin{equation} 
\label{dotuh}
\dot U_{\lambda_T^u \lambda_R^u}=h_{\lambda_T^u \lambda_R^u}(U,U^*)
\end{equation}
with the complex function
\begin{equation} 
\label{hbed}
h_{\lambda_T^u \lambda_R^u}(U,U^*)=\Lambda_{\lambda_T^u \lambda_R^u}U_{\lambda_T^u \lambda_R^u}+A_{\lambda_T^u \lambda_R^u} U_{\lambda_T^u \lambda_R^u}^2 
U_{-\lambda_T^u -\lambda_R^u}+B_{\lambda_T^u \lambda_R^u}U_{\lambda_T^u \lambda_R^u}U_{-\lambda_T^u \lambda_R^u}U_{\lambda_T^u -\lambda_R^u} \,.
\end{equation}
Here we have introduced the coefficients
\begin{eqnarray}
A_{\lambda_T^u \lambda_R^u}&=&-\frac{\gamma}{L_T L_R}\left(2-\frac{\gamma+\gamma^{2\lambda_T^u,2\lambda_R^u}}{2\Lambda_{\lambda_T^u \lambda_R^u}
-\Lambda_{2\lambda_T^u,2\lambda_R^u}}\right)\,,\label{1dA}\\
B_{\lambda_T^u \lambda_R^u}&=&-\frac{\gamma}{L_T L_R}\left[4-\frac{\gamma+
(\gamma^{2\lambda_T^u,0}-1)/2}{2\Lambda_{\lambda_T^u \lambda_R^u}-\Lambda_{2\lambda_T^u,0}}-\frac{\gamma+
(\gamma^{0,2\lambda_R^u}-1)/2}{2\Lambda_{\lambda_T^u \lambda_R^u}-\Lambda_{0,2\lambda_R^u}}\right] 
\label{1dB}
\end{eqnarray}
with the abbreviations $\gamma^{\lambda_T \lambda_R}:=f_{\lambda_T}^T f_{\lambda_R}^R$ and 
$\gamma:=\gamma^{\lambda_T^u,\lambda_R^u}=f_{\lambda_T^u}^T f_{\lambda_R^u}^R$.
Now we turn to the question whether the order parameter equations (\ref{dotuh})
represent a potential dynamics. 
To this end we derived in Ref.~\cite{thesis} a condition for the order 
parameter equations which allows one to conclude whether or not such a potential exists.
The potential criterion reads
\begin{equation} \label{bedkett}
\frac{\partial h_{\lambda_T^u \lambda_R^u}(U,U^*)}{\partial U_{\lambda_T^{u'} 
\lambda_R^{u'}}}=\frac{\partial h_{\lambda_T^{u'} 
\lambda_R^{u'}}^*(U,U^*)}{\partial U_{\lambda_T^u \lambda_R^u}^*}\,,
\end{equation}
which is, indeed, fulfilled for (\ref{hbed}). Furthermore, we derived 
in Ref.~\cite{thesis} the following conditions for determining 
the underlying potential:
\begin{equation}
\dot U_{\lambda_T^u \lambda_R^u}=-\frac{1}{2}\frac{\partial 
V(U,U^*)}{\partial U_{\lambda_T^u \lambda_R^u}^*} \,.
\label{Vu-mtmr}
\end{equation}
Integrating (\ref{Vu-mtmr}) yields the potential 
\begin{eqnarray}
V(U,U^*)&=&-2\Lambda_{\lambda_T^u \lambda_R^u}\left(U_{\lambda_T^u \lambda_R^u}
U_{-\lambda_T^u -\lambda_R^u}+U_{-\lambda_T^u \lambda_R^u}
U_{\lambda_T^u -\lambda_R^u}\right)\nonumber\\
 & &-A_{\lambda_T^u \lambda_R^u} \left(U_{\lambda_T^u \lambda_R^u}^2 
U_{-\lambda_T^u -\lambda_R^u}^2
+U_{-\lambda_T^u \lambda_R^u}^2 U_{\lambda_T^u -\lambda_R^u}^2\right)\nonumber\\
& &-2B_{\lambda_T^u \lambda_R^u} U_{\lambda_T^u \lambda_R^u}U_{-\lambda_T^u -\lambda_R^u}
U_{-\lambda_T^u \lambda_R^u}U_{\lambda_T^u -\lambda_R^u}\,.
\label{potkomplex}
\end{eqnarray}
\subsection{Real Order Parameters} \label{stringrop}

For technical purposes it has turned out to be useful to work with
complex order parameters so far. However, in order to investigate their contribution
to a one-to-one mapping between the strings, we have to transform them to 
real variables. We construct at first the real modes from the eigenfunctions 
(\ref{eigenf}), (\ref{saitenef}) of the linear 
operator $\hat L$ according to
\begin{eqnarray}
\hspace{-0.8cm}c_{\lambda_T \lambda_R}(t,r)&=&\frac{1}{2}\left[v_{\lambda_T \lambda_R}(t,r)+v_{-\lambda_T -\lambda_R}(t,r)\right]=
\frac{1}{\sqrt{L_T L_R}}\,\cos\left(\frac{2\pi}{L_T}\lambda_T t+\frac{2\pi}{L_R}\lambda_R r\right)
\,,\label{saitecreell}\\
\hspace{-0.8cm}s_{\lambda_T \lambda_R}(t,r)&=&-\frac{i}{2}\left[v_{\lambda_T \lambda_R}(t,r)-v_{-\lambda_T -\lambda_R}(t,r)\right]=
\frac{1}{\sqrt{L_T L_R}}\,\sin\left(\frac{2\pi}{L_T}\lambda_T t+\frac{2\pi}{L_R}\lambda_R r\right)
\,.\label{saitesreell}
\end{eqnarray}
These two modes span a real subspace. If we set
\begin{equation} \label{1drho}
a=\rho \cos\psi\,,\quad b=\rho \sin\psi\,;\quad \rho\geq 0\,, \quad \psi\in (-\pi,\pi]\,,
\end{equation}
the following relation results:
\begin{equation} 
\label{1dphase}
a\,c_{\lambda_T \lambda_R}(t,r)+b\,s_{\lambda_T \lambda_R}(t,r)=\rho \cos\left(\frac{2\pi}{L_T}\lambda_T t
+\frac{2\pi}{L_R}\lambda_R r-\psi\right)\,.
\end{equation}
Thus, the subspace consists of all phase-shifted functions of $\rho c_{\lambda_T \lambda_R}(t,r)$. 
Then the modes belonging to the unstable eigenvalue ($\lambda_T^u,\lambda_R^u$) are given by 
the modes $c_{\lambda_T^u \lambda_R^u}(t,r)$ and $c_{\lambda_T^u -\lambda_R^u}(t,r)$ as well as all 
phase-shifted functions. Rewriting the unstable part (\ref{UEXP})
\begin{eqnarray}
U(t,r)&=&U_{\lambda_T^u \lambda_R^u}v_{\lambda_T^u \lambda_R^u}(t,r)+U_{-\lambda_T^u 
-\lambda_R^u}v_{-\lambda_T^u -\lambda_R^u}(t,r)\nonumber\\
&&+U_{\lambda_T^u -\lambda_R^u}v_{\lambda_T^u -\lambda_R^u}(t,r)+U_{-\lambda_T^u \lambda_R^u}
v_{-\lambda_T^u \lambda_R^u}(t,r)
\end{eqnarray}
to real modes (\ref{saitecreell}), (\ref{saitesreell}), leads to
\begin{equation}
U(t,r)=u_1 c_{\lambda_T^u \lambda_R^u}(t,r)+u_2 s_{\lambda_T^u \lambda_R^u}(t,r)
+u_3 c_{\lambda_T^u -\lambda_R^u}(t,r)
+u_4 s_{\lambda_T^u -\lambda_R^u}(t,r)\,,\label{1d299}
\end{equation}
with real variables $u_j$:
\begin{eqnarray} 
\label{Uu}
U_{\lambda_T^u \lambda_R^u}&=&(u_1-iu_2)/2\,, \quad U_{-\lambda_T^u -\lambda_R^u}=(u_1+iu_2)/2\,,
\vspace{0.1cm}\nonumber\\
U_{\lambda_T^u -\lambda_R^u}&=&(u_3-iu_4)/2\,,\quad U_{-\lambda_T^u \lambda_R^u}=(u_3+iu_4)/2\,.
\end{eqnarray}
Inserting the transformations (\ref{Uu}) into 
the complex potential (\ref{potkomplex}), we obtain the following 
real potential
\begin{eqnarray}
V(u_i)&=&-\frac{\Lambda_{\lambda_T^u \lambda_R^u}}{2}(u_1^2+u_2^2+u_3^2+u_4^2)
-\frac{A_{\lambda_T^u \lambda_R^u}}{16} \left[\left(u_1^2+u_2^2\right)^2+\left(u_3^2+u_4^2\right)^2\right]\nonumber\\
&&-\frac{B_{\lambda_T^u \lambda_R^u}}{8} (u_1^2+u_2^2)(u_3^2+u_4^2)\,. \label{potreell1}
\end{eqnarray}
The corresponding 
equations of evolution for the real order parameters are determined from (\ref{potreell1}) according to
$\dot u_j=-\partial V(u_i)/\partial u_j\,.$
They read explicitly
\begin{eqnarray} \label{1du1-u4}
\dot u_1&=&\left[\Lambda_{\lambda_T^u \lambda_R^u}+\frac{A_{\lambda_T^u \lambda_R^u}}{4}(u_1^2+u_2^2)
+\frac{B_{\lambda_T^u \lambda_R^u}}{4}(u_3^2+u_4^2)\right]u_1\,,
\nonumber\\
\dot u_2&=&\left[\Lambda_{\lambda_T^u \lambda_R^u}+\frac{A_{\lambda_T^u \lambda_R^u}}{4}(u_1^2+u_2^2)
+\frac{B_{\lambda_T^u \lambda_R^u}}{4}(u_3^2+u_4^2)\right]u_2\,,
\nonumber\\
\dot u_3&=&\left[\Lambda_{\lambda_T^u \lambda_R^u}+\frac{A_{\lambda_T^u \lambda_R^u}}{4}(u_3^2+u_4^2)
+\frac{B_{\lambda_T^u \lambda_R^u}}{4}(u_1^2+u_2^2)\right]u_3\,,\nonumber\\
\dot u_4&=&\left[\Lambda_{\lambda_T^u \lambda_R^u}+\frac{A_{\lambda_T^u \lambda_R^u}}{4}(u_3^2+u_4^2)
+\frac{B_{\lambda_T^u \lambda_R^u}}{4}(u_1^2+u_2^2)\right]u_4\,.
\end{eqnarray}
\begin{figure}[t!]
\centerline{\includegraphics[scale=0.8]{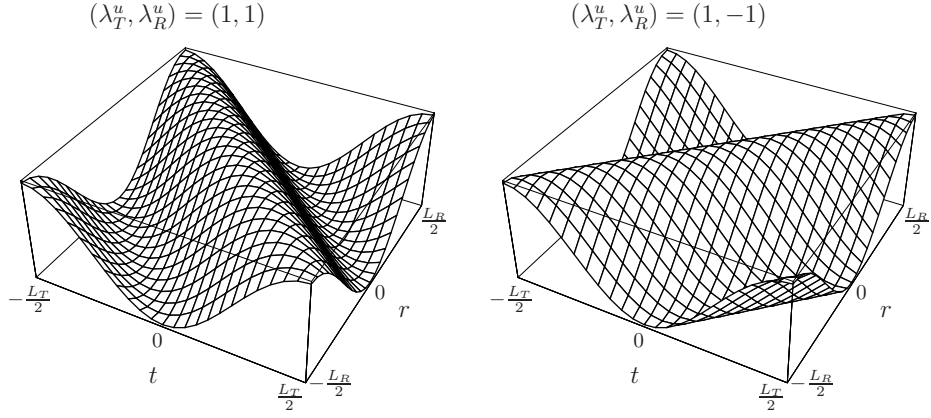}}
\caption{\label{11moden} \small Diagonal modes of different orientation according 
to (\ref{Usuperposition}) for the case $(\lambda_T^u,\lambda_R^u)=(1,1)$ and $(1,-1)$, respectively. 
Here the phase shifts are set to $\psi=\varphi=0$.}
\end{figure}
\subsection{Constant Phase Shift Angles} \label{konstantephasenwinkel}

According to Eq.~(\ref{1dphase}) the unstable part (\ref{1d299})
can be written as a superposition of two diagonal modes of different orientation 
\begin{equation} \label{Usuperposition}
U(t,r)=\xi\cos\left[\frac{2\pi}{L_T}\lambda_T^u t+\frac{2\pi}{L_R}\lambda_R^u r-\psi\right]
+\eta\cos\left[\frac{2\pi}{L_T}\lambda_T^u t-\frac{2\pi}{L_R}\lambda_R^u r-\varphi\right]
\end{equation}
as is illustrated in Fig.~\ref{11moden}.
With (\ref{1drho}) and (\ref{1dphase}) we have
$u_1=\xi\cos\psi\,,u_2=\xi\sin\psi\,,u_3=\eta\cos\varphi\,,u_4=\eta\sin\varphi\,.$
Then the amplitudes of the phase-shift diagonal modes read
\begin{equation}
\xi=\sqrt{u_1^2+u_2^2}\,, \qquad 
\eta=\sqrt{u_3^2+u_4^2}
\label{1dtrafoeta}
\end{equation}
and the phase angles are given by $\tan\psi=u_1/u_2\,,\,\tan\varphi=u_3/u_4\,.$
From the order parameter equations (\ref{1du1-u4}) it follows
$\dot u_1/\dot u_2=u_1/u_2\,,\,\dot u_3/\dot u_4=u_3/u_4\,.$
Thus, performing a separation of variables and a subsequent integration leads to the relation
$u_1/u_2={\rm const}\,,\,u_3/u_4={\rm const}\,.$
Consequently, the four real equations (\ref{1du1-u4}) are reduced to 
two equations for the mode amplitudes $\xi$ and $\eta$:
\begin{eqnarray} \label{saitexieta}
\dot\xi&=&\left(\Lambda_{\lambda_T^u \lambda_R^u}+\frac{A_{\lambda_T^u \lambda_R^u}}{4}\xi^2
+\frac{B_{\lambda_T^u \lambda_R^u}}{4}\eta^2\right)\xi\,,\nonumber\\
\dot\eta&=&\left(\Lambda_{\lambda_T^u \lambda_R^u}+\frac{A_{\lambda_T^u \lambda_R^u}}{4}\eta^2
+\frac{B_{\lambda_T^u \lambda_R^u}}{4}\xi^2\right)\eta\,.
\end{eqnarray}
The corresponding potential is
\begin{equation} \label{saiteallgpot}
V(\xi,\eta)=-\frac{\Lambda_{\lambda_T^u \lambda_R^u}}{2}(\xi^2+\eta^2)-\frac{A_{\lambda_T^u \lambda_R^u}}{16}(\xi^4+\eta^4)
-\frac{B_{\lambda_T^u \lambda_R^u}}{8}\xi^2\eta^2\,.
\end{equation}
Thus, we have reduced the four complex order parameter equations (\ref{dotuh}), (\ref{hbed}) 
to two real order parameter equations (\ref{saitexieta}) with the 
potential (\ref{saiteallgpot}). 
\subsection{Monotonous Cooperativity Functions} \label{lambda11fall}

So far our considerations are valid for arbitrary unstable modes $(\lambda_T^u,\lambda_R^u)\,.$ 
According to the eigenvalue spectrum (\ref{ewsaite}) the unstable modes are determined 
by the expansion coefficients $f_{\lambda}$ of the cooperativity functions. 
We therefore derive in this subsection some basic properties of
these coefficients. In particular, we investigate the consequences of monotonically 
decreasing cooperativity functions for their expansion coefficients $f_{\lambda}$.
As $c(x)$ is positive and normalized, we conclude $|f_{\lambda}|\leq1$.
Using the Euler formula, the symmetry $c(x)=c(-x)$, and integrating by parts, the expansion coefficients 
can be written in the form
\begin{equation}
f_{\lambda}=-\frac{L}{\pi \lambda}\int\limits_0^{L/2}c\,'(x)\sin\left(\frac{2\pi}{L}\lambda x\right)\,dx\,,
\end{equation}
which makes the symmetry $f_{\lambda}=f_{-\lambda}$ manifest.
If we assume monotonically decreasing cooperativity functions, i.e.~$dc/dx < 0$ for 
$x\in [0,L/2]$, we obtain $f_1>0$.
Furthermore, we can show that $f_1$ is the largest expansion coefficient by
considering the expression
\begin{equation} 
\label{1dfdiff}
f_1-f_{\lambda}=-\frac{L}{\pi}\int\limits_0^{L/2}c\,'(x)\left[\sin\left(\frac{2\pi}{L}x\right)
-\frac{1}{\lambda}\sin\left(\frac{2\pi}{L}\lambda x\right)\right]\,dx\,.
\end{equation}
Because of $c\,'(x)<0$ it follows indeed
$f_1-f_{\lambda}>0\quad\forall\,\lambda\not=0,\pm 1$. Together
with $|f_{\lambda}|<1$ the maximum eigenvalue of (\ref{ewsaite}) 
results to be $\Lambda_{\rm max}=-\alpha+f_1^T f_1^R$.
Hence in this case there are four unstable modes $(\lambda_T^u,\lambda_R^u)=(\pm 1,\pm 1)$, which 
corresponds to the result obtained in Ref.~\cite{Malsburg}. 
However, the most fundamental insight of our more general analysis is 
that the real order parameter equations (\ref{saitexieta})
are also valid in the case where the cooperativity functions are not monotonic so that
any mode $(\lambda_T^u,\lambda_R^u)$ can become unstable.
It is plausible that there is a pathological development in animals which corresponds
to this case.
\begin{figure}[t!]
\centerline{\includegraphics[scale=0.9]{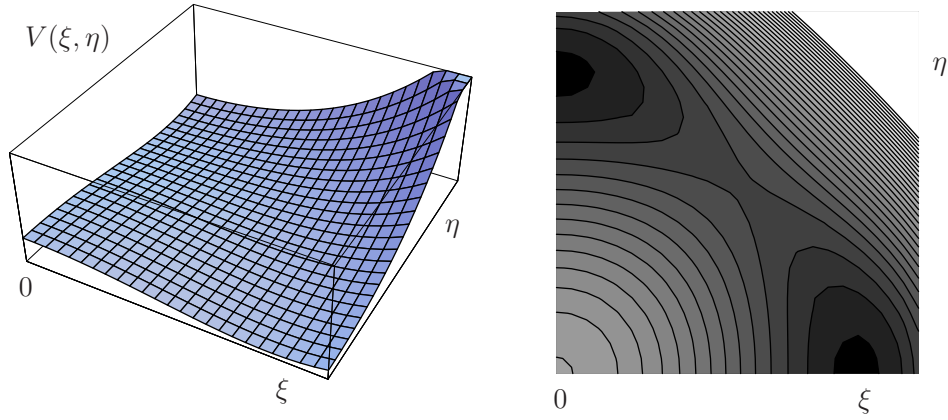}}
\caption{\label{potbild} \small The potential $V(\xi,\eta)$ according to Eq.~(\ref{pot11fall}) 
with $\Lambda>0$. The originally stable state $\xi=\eta=0$ becomes unstable. The system settles into 
one of the two minima, i.e.~one of the two modes vanishes. The right plot shows the 
equipotential lines. Dark grey values correspond to small values of the potential $V$.}
\end{figure}

\subsection{Potential Properties} \label{saitenpoteigen}

We now analyze the properties of the potential (\ref{saiteallgpot}). For this purpose we restrict
ourselves from now on to the unstable modes $(\lambda_T^u,\lambda_R^u)=(\pm 1,\pm 1)$, whose 
indices will be discarded for the sake of simplicity. Then the potential (\ref{saiteallgpot}) 
reads
\begin{equation} \label{pot11fall}
V(\xi,\eta)=-\frac{\Lambda}{2}(\xi^2+\eta^2)-\frac{A}{16}(\xi^4+\eta^4)-\frac{B}{8}\xi^2\eta^2\,,
\end{equation}
where the coefficients $A$, $B$ follow from (\ref{1dA}), (\ref{1dB}) to be
\begin{eqnarray}
A&=&-\frac{\gamma}{L_T L_R}\left(2-\frac{\gamma+\gamma^{2,2}}{2\Lambda-\Lambda_{2,2}}\right)\,, \label{saiteA11}\\
B&=&-\frac{\gamma}{L_T L_R}\left[4-\frac{\gamma+(\gamma^{2,0}-1)/2}{2\Lambda-\Lambda_{2,0}}
-\frac{\gamma+(\gamma^{0,2}-1)/2}{2\Lambda-\Lambda_{0,2}}\right]\,. 
\label{saiteB11}
\end{eqnarray}
From the condition $\nabla V=0$ we determine the extrema of $V(\xi,\eta)$ and assign 
them to a minimum, a maximum, or a saddle point.
In the unstable region with $\Lambda>0$ 
the potential $V(\xi,\eta)$, which is depicted in Fig.~\ref{potbild}, has
\begin{itemize}
\item a relative maximum at $\,P_1(0,0)\,,$\\
\item two relative minima at $\,P_2(0,\sqrt{-4\Lambda/A})$ and $P_3(\sqrt{-4\Lambda/A},0)\,,$\\
\item a saddle point at $\,P_4(\sqrt{-4\Lambda/(A+B)},\sqrt{-4\Lambda/(A+B)})\,.$
\end{itemize}
In the stable region with $\Lambda<0$ only the relative minimum $\xi=\eta=0$ does exist. Initially, the 
system is in the stable uniform state $w_0(t,r)=1$. This state becomes unstable if the 
control parameter $\alpha$ is decreased to the critical value $\alpha_c=f_1^T f_1^R$. The 
eigenvalue $\Lambda_{\rm max}=-\alpha+f_1^T f_1^R$ becomes positive, and the minimum passes into a maximum. 
The system settles into one of the two equivalent minima, i.e.~a symmetry breaking takes 
place. Thereby the two modes compete with each other and, subsequently, one of the two modes 
vanishes. Which of them vanishes depends on the initial 
conditions of $\xi$ and $\eta$. If the condition $\eta(0)>\xi(0)$ is fulfilled, the $\xi$-mode vanishes, 
and vice versa.
\subsection{One-To-One Retinotopy} \label{stringotor}
\begin{figure}[t!]
\centerline{\includegraphics[scale=0.8]{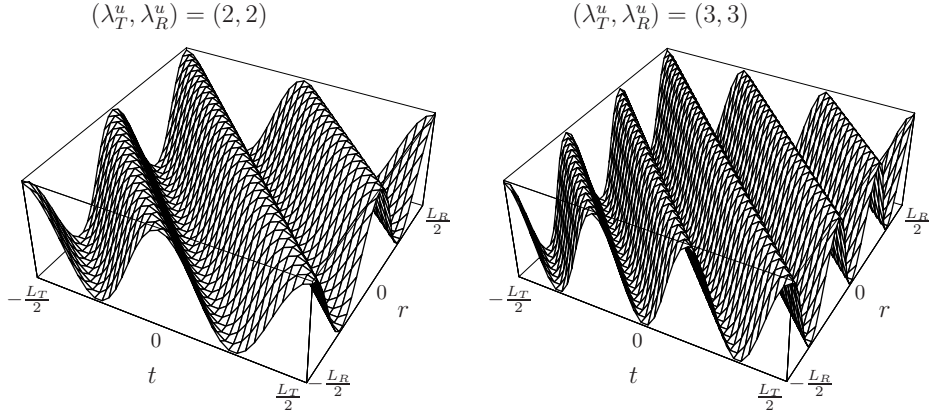}}
\caption{\label{diagmoden} \small Higher diagonal modes, which are excited by 
the unstable mode $(\lambda_T^u,\lambda_R^u)=(1,1)$ of Fig.~\ref{11moden}. They amplify the 
retinotopic character of the $(1,1)$-mode.}
\end{figure}

In the following we assume that, according to the potential dynamics discussed above, only 
one of the two modes remains. These two modes show a pronounced maximum for $t=-r$ and $t=r$, 
respectively, as is shown in Fig.~\ref{11moden}. 
To assess the influence of higher modes, we calculate 
the center manifold $S(U)$ for the case $\xi=0$ and $\eta\not=0$ and set 
$u_4=0$ without loss of generality. Then it follows from (\ref{1dtrafoeta}) that $\eta=u_3$, 
and we obtain for the unstable part (\ref{1d299})
\begin{equation}
U(t,r)=\eta\cos\left(\frac{2\pi}{L_T}t-\frac{2\pi}{L_R}r\right)\,.
\end{equation}
With the center manifold (\ref{1dzentrmann}) the stable part (\ref{SEXP}), (\ref{HN}) reads explicitly
\begin{equation}
S(U)=\frac{2 \gamma}{\sqrt{L_T L_R}\left(2\Lambda-\Lambda_{2,2}\right)}\,\eta^2 \cos\left(\frac{4\pi}{L_T}t
-\frac{4\pi}{L_R}r\right)\,.
\end{equation}
Thus, those modes are excited which strengthen the retinotopic character of the projection. 
With the help of the complex modes it can be seen that this is also the case for higher modes, 
i.e.~for $(\lambda_T^u,\lambda_R^u)=(1,1)$ exclusively the modes $(2,2)$, $(3,3)$ etc. are excited, 
which are depicted in Fig.~\ref{diagmoden}. 
Therefore, we follow Ref.~\cite{Malsburg} and use an ansatz which contains 
only diagonal modes and insert it into the H{\"a}ussler-von der Malsburg equations (\ref{HAUS}).
If we restrict ourselves to special cooperativity functions, 
the resulting recursion relations can be solved analytically by using the method of 
generating function.
Note that our derivation of the solution of the recursion relations corresponds to the gravitating chain in Ref.~\cite{Malsburg}.
\subsubsection{Recursion Relations}
Motivated by the above remarks we investigate the H{\"a}ussler-von der Malsburg equations for strings with
the ansatz
\begin{equation} 
\label{1dansatz}
w(t,r)=\sqrt{L_T L_R}\sum_{\lambda=-\infty}^{\infty}w_{\lambda} v_{\lambda,-\lambda}(t,r)\,,
\end{equation}
where $v_{\lambda,-\lambda}(t,r)$ is defined by (\ref{eigenf}) and (\ref{saitenef}). 
Thus, taking into account the decomposition (\ref{1dct}) of the
cooperativity functions, 
the H{\"a}ussler-von der Malsburg equations (\ref{HAUS}) can be written as
\begin{equation} \label{hsler}
\dot w(t,r)=-\alpha[w(t,r)-1]+w(t,r)\sqrt{L_T L_R}\sum_{\lambda=-\infty}^{\infty} w_{\lambda} f_{\lambda}^T f_{\lambda}^R v_{\lambda,-\lambda}(t,r)-p(w)\,w(t,r)\,,
\end{equation}
where we have introduced the abbreviation
\begin{equation} 
\label{defpvonw}
p(w)=\sum_{\lambda=-\infty}^{\infty} w_{-j}w_j f_j^T f_j^R\,.
\end{equation}
Inserting the ansatz (\ref{1dansatz}) into (\ref{hsler}) 
and comparing the coefficients of the linearely independent functions $v_{\lambda,-\lambda}(t,r)$ yields 
\begin{eqnarray}
\dot w_0&=&-\left[\alpha+p(w)\right](w_0-1)\,,
\label{1dretinogl1}\\
\dot w_{\lambda}&=&-\left[\alpha+p(w)\right]w_{\lambda}+
\sum_{j=-\infty}^{\infty}w_{\lambda-j}w_j f_j^T f_j^R\,,\quad \lambda\not=0\,. 
\label{1dretinogl}
\end{eqnarray}
As $w(t,r)$ is positive \cite{gpw1}, 
we obtain that $p(w)>0$ and $\alpha+p(w)>0$.
Therefore, the stationary state is determined from (\ref{1dretinogl1}) to be
$w_0=1$.
\subsubsection{Special Cooperativity Functions}
We restrict our further considerations to the following form 
of the cooperativity functions (\ref{1dct}):
$f_0=1\,,f_1\not=0\,,f_j=0\, \mbox{ for } j\not=0,\pm 1$.
With the abbreviation $\gamma:=f_1^T f_1^R$ the previous result (\ref{defpvonw}) can be written as
$p(w)=1+2\gamma w_1 w_{-1}$,
so that the equations (\ref{1dretinogl}) for the stationary case reduce to
the recursion relation
\begin{equation}
(\alpha+2\gamma w_1^2)w_{\lambda}=\gamma w_1(w_{\lambda-1}+w_{\lambda+1})\,,\quad \lambda\not=0\,.
\label{saiteugl}
\end{equation}
\subsubsection{Generating Function}
To solve the recursion relation (\ref{saiteugl}), 
we define the generating function
\begin{equation} 
\label{saite8128}
E(z)=\sum_{\lambda=-\infty}^{\infty}w_{\lambda}z^{\lambda}\,.
\end{equation}
Multiplying (\ref{saiteugl}) with $z^{\lambda}+z^{-\lambda}$ and performing the sum 
from $\lambda=1$ up to infinity yields
to a linear algebraic equation which is solved by
\begin{equation} 
\label{saite8130}
E(z)=\frac{\alpha}{\alpha+2\gamma w_1^2-\gamma w_1\left(z+z^{-1}\right)}\,.
\end{equation}
To determine the coefficients $w_{\lambda}$ we expand the generating function 
(\ref{saite8130}) into a Taylor series: 
\begin{equation}
\label{E}
E(z)=-\frac{\alpha}{\left(\alpha+2\gamma w_1^2\right)w\left(z_1-z_1^{-1}\right)}\,
\left[\sum_{\lambda=1}^{\infty}z_1^{\lambda}(z^{\lambda}+z^{-\lambda})+1\right]\,,\quad|z_1|<|z|<|z_1|^{-1}
\end{equation}
with the abbreviations
\begin{equation} \label{saitez12}
w=\frac{\gamma w_1}{\alpha+2\gamma w_1^2}\,,\qquad z_1=\frac{1}{2w}(1+\sqrt{1-4w^2}\,)\,.
\end{equation}
Comparing (\ref{E}) with (\ref{saite8128}) by taking into account $w_0=1$
determines $w_1$ to be
\begin{equation}
\label{U1}
w_1=\sqrt{\frac{\gamma-\alpha}{\gamma}}\,.
\end{equation}
Thus, together with (\ref{saitez12}) it follows $z_1=w_1$, and the 
remaining coefficients turn out to be
$w_{\lambda}=w_1^{|\lambda|}$,
which is valid not only for $\lambda\not=0$ but also for $\lambda=0$ due to $w_0=1$.
\begin{figure}[t!]
\centerline{\includegraphics[scale=0.8]{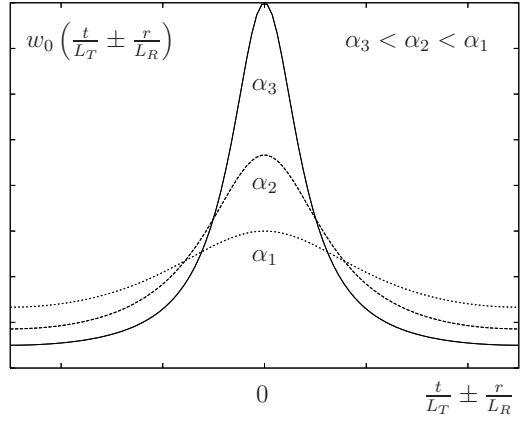}}
\caption{\label{deltapeak} \small Decreasing the control parameter $\alpha$ 
to smaller values, we read off from (\ref{U1}) and (\ref{w0872}) that the connection 
weight converges to Dirac's delta function (\ref{w0878}).}
\end{figure}
\subsubsection{Limiting Cases}
By inserting the latter result into (\ref{1dansatz}) we obtain 
\begin{equation}
w_0(t,r)=\frac{1-w_1^2}{1-2 w_1 \cos\left(\frac{2\pi}{L_T}t
-\frac{2\pi}{L_R}r\right)+w_1^2}\,. \label{w0872}
\end{equation}
For $w_1=0$ the stationary uniform state reduces to
$w_0(t,r)=1\, \forall t\in[0,L_T)\,,\, r\in[0,L_R)$.
In the case $w_1=1$, i.e.~$w_{\lambda}=1\,\,\forall \lambda$, 
we find with the help of the Poisson formula \cite{klein}
\begin{equation} \label{w0878}
w_0(t,r)=\delta\left(\frac{t}{L_T}-\frac{r}{L_R}\right)\,.
\end{equation}
Hence we have a situation which is illustrated in Fig.~\ref{deltapeak}: If the control 
parameter $\alpha$ is in the neighborhood of $\gamma$, 
the connection weight is essentially uniform with a small maximum 
for $t/L_T=\pm r/L_R$. Further decreasing of $\alpha$ leads to a sharpening of the projection. 
In the case $\alpha\to 0$ the projection becomes Dirac's delta function, i.e.~a perfect 
one-to-one retinotopy is achieved. This means that the undifferentiated growth of new 
synaptic contacts comes to an end when the ordered projection between retina and tectum 
is fully developed.
\subsection{Comparison with Linear Chains} \label{stringclc}
Finally, we compare our results for strings with those of
Ref.~\cite{Malsburg} where retina and tectum were treated as linear chains consisting 
of $N$ cells, respectively. In that reference, the order parameter equations read
\begin{equation} \label{1dzij}
\dot U_{ij}=[\Lambda-\gamma(2-a)U_{ij}U_{-i-j}+(4-b'-b'')U_{i-j}U_{-ij}]U_{ij}
\end{equation}
with the abbreviations
\begin{equation}
a=-\frac{\gamma+\gamma^{2,2}}{\Lambda_{22}}\,,\quad
b'=-\frac{\gamma+(\gamma^{2,0}-1)/2}{\Lambda_{20}}\,,\quad
b''=-\frac{\gamma+(\gamma^{0,2}-1)/2}{\Lambda_{02}}\,.
\end{equation}
The comparison of the coefficients $A$, $B$ according to (\ref{saiteA11}), 
(\ref{saiteB11}) with $\gamma(2-a)$, $\gamma(4-b'-b'')$ exhibits two differences: the factor 
$1/L_T L_R$ in $A$ and $B$ as well as the term $2\Lambda$ in the denominator. The absence 
of the 
\begin{figure}[t!]
\centerline{\includegraphics[scale=0.7]{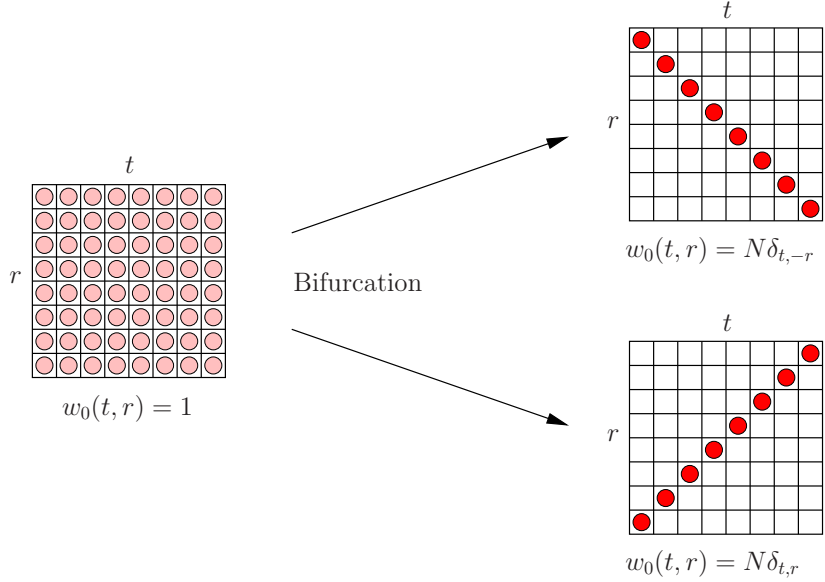}}
\caption{\label{bifurkation} \small Bifurcation in the vicinity of the instability 
point for the linear chain as analyzed in Ref.~\cite{Malsburg}. The quadratic arrangement of 
the two linear chains allows a concise representation of the connection weights. 
Dark gray means high connection weights between the corresponding cells $t$ and $r$. 
At the uniform initial state all connection weights are equal. The bifurcation drives 
the system into one of the two possible states, which differ in their orientation. 
Decreasing the control parameter $\alpha$ to zero leads to a 
one-to-one retinotopy. Instead of a delta function in the continuous case, here the 
retinotopic order is described by Kronecker deltas.}
\end{figure}
corresponding factor $1/N^2$ in Ref.~\cite{Malsburg} stems from the circumstance 
that the eigenfunctions were not normalized there. Physically 
more interesting is the appearance of 
the term $2\Lambda$ in the denominator of $A$, $B$. The reason for this is that we have used 
the mathematically correct equation for determining the center manifold (\ref{HNN}) 
according to Refs.~\cite{gpw1,wwp}, whereas in Ref.~\cite{Malsburg} the center manifold is adiabatically approximated by $\dot S=0$. 
However, this ad-hoc method for implementing 
the adiabatic approximation, which is frequently used in the 
literature, is only justified for real eigenvalues. As the eigenvalues of the strings
are real, we deduce for the vicinity of the instability point the relation
$\Lambda=-\alpha+\gamma\approx 0$.
Thus, the coefficients (\ref{saiteA11}), (\ref{saiteB11}) turn 
into those of Ref.~\cite{Malsburg} and the adiabatic 
approximation $\dot S=0$ can be applied here.\\

Furthermore, our results for the continuous case are analogous to the results for discrete cell 
arrays. Also the transition to a perfect one-to-one retinotopy 
takes place in a corresponding way, as is illustrated in Fig.~\ref{bifurkation}. 
Thus, we conclude that
our geometry-independent model for the emergence of retinotopic projections 
developed in Ref.~\cite{gpw1} contains as a special case the results of Ref.~\cite{Malsburg}. 
In addition, we have extended the range of validity, 
i.e.~the domain around the instability with $\Lambda=0$, 
where the order parameter equations represent a quantitatively good approximation, as 
we have derived a more 
precise form of the center manifold (\ref{HNN}). 
\section{Planes} \label{planes}

In this section we extend our discussion to two dimensions where
the cell sheets are assumed to be planes of side lengths 
$L_1^T,L_2^T$ and $L_1^R,L_2^R$, respectively. To obtain a consistent solution we 
assume again periodic boundary conditions, i.e.~the 
cell sheets are modelled as surfaces of tori. 
We start with presenting the linear analysis in Subsecs.~\ref{planeef} and \ref{planeip}.
Afterwards, in Subsecs.~\ref{planeope} and \ref{planerv} it turns out that we have to calculate
in total sixteen order parameter equations where the 
quadratic term vanishes, as in the case of strings, and where again
selection rules reduce the number of cubic terms. This order parameter
dynamics turns out to be complicated as the two dimensions do not decouple in
a trivial way. Therefore, we have to restrict our analytical discussion of the
order parameter dynamics to physiologically interesting special cases. If we set all modes to zero except for one, we
find retinotopy only in one dimension, which is shown in Subsec.~\ref{planerp}. In a next step we consider the 
superposition of two retinotopic modes and investigate the necessary conditions for
their coexistence. Such a situation occurs, for instance, when the cooperativity
function of the tectum is monotonically decreasing, whereas the cooperativity function
of the retina is not monotonic. In Subsec.~\ref{planecm} we show that taking into account the
center manifold contribution or higher modes leads to a sharpening of the retinotopic
character of the projection between planar retina and tectum.\\
\subsection{Eigenfunctions} \label{planeef}
In the following we consider both retina and tectum to be planes
with side lenghts $L_1$ and $L_2$. 
The points on the plane are represented by Cartesian coordinates $x=(x_1,x_2)\,,x_1\in[0,L_1)\,,\,x_2\in[0,L_2)$.
The magnitude of the plane is given by $M=L_1 L_2$.  
The corresponding eigenvalue equation
$\Delta \psi(x)=\chi \psi(x)$
is solved for periodic boundary conditions, 
i.e.~$\psi_j(x_j)=\psi_j(x_j+L_j)$ by the complete 
orthonormal system of eigenfunctions
\begin{equation} 
\label{ebef}
\psi_{\lambda}(x)=\frac{1}{\sqrt{L_1 L_2}}\exp\left[2 \pi i 
\left( \frac{\lambda_1 x_1}{L_1} - \frac{\lambda_2 x_2}{L_2}\right)\right]\,,
\end{equation}
where $\lambda=(\lambda_1,\lambda_2)$. The cooperativity function $c(x-x')$ is expanded according to (\ref{CEX}) in this basis:
\begin{eqnarray} \label{ebcoopexp}
c(x-x')= \frac{1}{L_1 L_2}\,\sum_{\lambda_1,\lambda_2} f_{(\lambda_1,\lambda_2)} \exp \left\{
2 \pi i \left[ \frac{\lambda_1 (x_1 - x_1')}{L_1} - \frac{\lambda_2 (x_2 - x_2')}{L_2} \right]
\right\} \, .
\end{eqnarray}
Note that again the expansion coefficients $f_{(\lambda_1,\lambda_2)}$ of the 
cooperativity functions are independent of the signs of the parameters $\lambda_1, \lambda_2$, as
the cooperativity functions should be symmetric with 
respect to their arguments: $c(x-x')=c(x'-x)$.
This requirement and the linear independence of the 
exponential functions leads to
$f_{(\lambda_{1},\lambda_{2})}=f_{(\pm \lambda_{1},\pm \lambda_{2})}$.
From now on we assume that the cooperativity functions 
decouple with respect to the two dimensions:
$c(x-x')=c_1(x_1-x_1')c_2(x_2-x_2')$.
As the individual cooperativity functions can be expanded according to
\begin{equation}
c_j(x_j-x_j')=\frac{1}{L_j}\sum_{\lambda_j}f_{\lambda_j} 
\exp\left[i\frac{2\pi}{L_j} \lambda_j (x_j-x_j')\right]\,,\quad j=1,2\,,
\end{equation}
the decoupling amounts to a factorization of the expansion coefficients:
$f_{(\lambda_1,\lambda_2)}=f_{\lambda_1}f_{\lambda_2}$.
With this we allow for both 
isotropic and certain anisotropic cooperativity functions. This is an interesting 
feature as it is reasonable to assume that real cell sheets have a preferential 
direction. 
\subsection{Instability Point} \label{planeip}
We analyze which modes become unstable. According to (\ref{ewsaite}) 
this depends on the expansion coefficients and the maximum eigenvalues are given by
$\Lambda_{\lambda_T^u \lambda_R^u}=-\alpha+f_{\lambda_T^u}^T f_{\lambda_R^u}^R$.
By doing so we require that all unstable modes become unstable {\it simultaneously}. 
This requirement is due to the fact that the order parameter equations should be approximately valid  
in the vicinity of the instability point. If the eigenvalues of the corresponding 
unstable modes would differ significantly, the situation that all modes are in 
the unstable region would have the consequence that the maximum eigenvalue would be larger than zero, i.e.~far 
away from the instability point. Thus, the order parameter equations would be no 
adequate approximation. Consequently, we only consider the case that
$\Lambda_{\lambda_T^u \lambda_R^u}=\Lambda_{\lambda_T^{u'} \lambda_R^{u'}}\,\forall \lambda_T^u,\lambda_R^u,\lambda_T^{u'},\lambda_R^{u'}\,,$
from which follows
$f_{\lambda_T^u}^T=f_{\lambda_T^{u'}}^T\,,f_{\lambda_R^u}^R=f_{\lambda_R^{u'}}^R 
\,\forall \lambda_T^u,\lambda_R^u,\lambda_T^{u'},\lambda_R^{u'}\,.$
However, it is possible that $f_{\lambda_T^u}^T\not=f_{\lambda_R^u}^R$.
From now on the unstable modes are assumed to be given by 
\begin{equation} \label{eb1010inst}
\lambda^u=(1,0),(-1,0),(0,1),(0,-1)\,.
\end{equation}
This occurs, for instance, for monotonically decreasing cooperativity functions where we obtain the relation 
$f_1^T>f_{\lambda}^T$ ($\lambda\not=0,\pm 1$), by analogy with strings (see Sec.~\ref{lambda11fall}). Then the maximum 
expansion coefficient is given by $f_{\lambda^u}=f_{(\lambda_1^u,\lambda_2^u)}=f_{\lambda_1^u} 
f_{\lambda_2^u}$ for $\lambda_1^u=0,\lambda_2^u=\pm 1$ and $\lambda_1^u=\pm 1,\lambda_2^u=0$, respectively. 
Note, however, that the unstable modes (\ref{eb1010inst}) could also arise
for non-monotonic cooperativity functions as we will see below.
\subsection{Order Parameter Equations} \label{planeope}
We specialize the order parameter equations (\ref{OPE}) to planes. 
At first we determine the integrals (\ref{abkurzI}), (\ref{abkurzJ}) of products of eigenfunctions (\ref{ebef}), which read
\begin{eqnarray}
I_{\lambda^{(1)} \lambda^{(2)}\ldots\, \lambda^{(n)}}^{\lambda}&=&\left(\frac{1}{L_1 L_2}\right)^{(n-1)/2}\delta_{\lambda^{(1)}+\lambda^{(2)}+\ldots+\lambda^{(n)},\lambda}\,,\label{ebI}\\
J_{\lambda^{(1)} \lambda^{(2)}\ldots\, \lambda^{(n)}}&=&\left(\frac{1}{L_1 L_2}\right)^{(n-2)/2}\delta_{\lambda^{(1)}+\lambda^{(2)}+\ldots+\lambda^{(n)},0}\,.
\label{ebJ}
\end{eqnarray}
For the order parameter equations (\ref{OPE}) we need in (\ref{AA}), (\ref{BB}), (\ref{HNN}) the special cases 
\begin{equation} \label{ebintegrale}
I_{\lambda' \, \lambda''}^{\lambda}=\frac{1}{\sqrt{L_1 L_2}}\,\delta_{\lambda'+\lambda'',\lambda}\,, 
\quad I_{\lambda' \, \lambda'' \, \lambda'''}^{\lambda}=\frac{1}{L_1 L_2}\,\delta_{\lambda'+\lambda''+\lambda''',\lambda}\,,
\quad J_{\lambda' \, \lambda''}=\delta_{\lambda',-\lambda''}\,.
\end{equation}
Note that the unstable modes (\ref{eb1010inst}) have the property $\lambda^{u'}+\lambda^{u''}\not=\lambda^u$. 
Thus, the quadratic coefficient (\ref{AA}) vanishes due to the first integral of (\ref{ebintegrale}):
$A^{\lambda_T^u, \lambda_T^u{}' \lambda_T^u{}''}_{\lambda_R^u, \lambda_R^u{}' \lambda_R^u{}''}  = 0 $.
To yield a concise calculation of the order parameter 
equations, we introduce modes $\bar\lambda^u$ which are complementary to the unstable modes $\lambda^u$ by permuting the two components: 
\begin{equation}
\bar\lambda^u=\left\{\begin{array}{cc}
(0,\pm 1)& \mbox{ if }\,\lambda^u=(\pm 1,0)\,,\\
(\pm 1,0)& \mbox{ if }\,\lambda^u=(0,\pm 1)\,.
\end{array} \right.
\end{equation}
The integrals (\ref{ebintegrale}) involve selection rules for the appearance of cubic terms (\ref{BB}) which are analogous to those for strings. This condition 
turns out to be $\lambda^{u'}+\lambda^{u''}+\lambda^{u'''}=\lambda^{u}$ for (\ref{eb1010inst}),
which leads to the following nine possibilities:
\begin{eqnarray}
(\lambda^{u'},\lambda^{u''},\lambda^{u'''})&=&(\lambda^u,\lambda^u,-\lambda^u),(\lambda^u,-\lambda^u,\lambda^u),(-\lambda^u,\lambda^u,\lambda^u),(\lambda^u,\bar\lambda^u,-\bar\lambda^u),(\bar\lambda^u,\lambda^u,-\bar\lambda^u),\nonumber\\
 & &(\bar\lambda^u,-\bar\lambda^u,\lambda^u),(\lambda^u,-\bar\lambda^u,\bar\lambda^u),(-\bar\lambda^u,\lambda^u,\bar\lambda^u),(-\bar\lambda^u,\bar\lambda^u,\lambda^u)\,.
\end{eqnarray}
In this way it can be shown that in total 14 possible cubic terms have to be taken into account.
The resulting order parameter equations read
\begin{eqnarray}
&&\hspace{-1.1cm}\dot U_{\lambda_T^u \lambda_R^u}=c_1 (U_{\lambda_T^u \lambda_R^u})^2 U_{-\lambda_T^u -\lambda_R^u}+c_2 U_{\lambda_T^u \lambda_R^u}U_{-\lambda_T^u \lambda_R^u}U_{\lambda_T^u -\lambda_R^u}+c_3 U_{\lambda_T^u \lambda_R^u}\left(U_{\bar\lambda_T^u \lambda_R^u}U_{-\bar\lambda_T^u -\lambda_R^u}\right.\nonumber\\
&&\hspace{-1.0cm}\left.+U_{\bar\lambda_T^u -\lambda_R^u}U_{-\bar\lambda_T^u \lambda_R^u}\right)+c_4 U_{\bar\lambda_T^u \lambda_R^u}U_{\lambda_T^u -\lambda_R^u}U_{-\bar\lambda_T^u \lambda_R^u}+c_5 U_{\lambda_T^u \lambda_R^u}\left(U_{\lambda_T^u \bar\lambda_R^u} U_{-\lambda_T^u -\bar\lambda_R^u}\right.\left.+U_{-\lambda_T^u \bar\lambda_R^u}U_{\lambda_T^u -\bar\lambda_R^u}\right)\nonumber\\
 & &\hspace{-1.0cm}+c_6 U_{-\lambda_T^u \lambda_R^u} U_{\lambda_T^u \bar\lambda_R^u}U_{\lambda_T^u -\bar\lambda_R^u}+c_7 U_{\lambda_T^u \lambda_R^u}\left(U_{\bar\lambda_T^u \bar\lambda_R^u}U_{-\bar \lambda_T^u -\bar \lambda_R^u}+U_{\bar\lambda_T^u -\bar\lambda_R^u} U_{-\bar\lambda_T^u \bar\lambda_R^u}\right)\nonumber\\
& &\hspace{-1.0cm}+c_8 \left(U_{\bar\lambda_T^u \lambda_R^u} U_{\lambda_T^u \bar\lambda_R^u} U_{-\bar\lambda_T^u -\bar\lambda_R^u}+U_{-\bar\lambda_T^u \lambda_R^u} U_{\lambda_T^u \bar\lambda_R^u} U_{\bar\lambda_T^u -\bar\lambda_R^u}\right.\nonumber\\
 & &\hspace{-1.0cm}\left.+U_{\bar\lambda_T^u \lambda_R^u} U_{\lambda_T^u -\bar\lambda_R^u} U_{-\bar\lambda_T^u \bar\lambda_R^u}+U_{-\bar\lambda_T^u \lambda_R^u} U_{\lambda_T^u -\bar\lambda_R^u} U_{\bar\lambda_T^u \bar\lambda_R^u}\right)\,.\label{ebopgkomplex}
\end{eqnarray}
The coefficients $c_1$--$c_8$ are fully determined by the expansion coefficients
$f_{\lambda}$ of the cooperativity functions and the control parameter $\alpha$, as is documented in Ref.~\cite{thesis}.

\subsection{Real Variables} \label{planerv}

To investigate how the complex order parameters contribute to the one-to-one-retinotopy between the planes, we have to transform them to real variables. To this end we introduce the transformation 
\begin{equation} \label{ebinvtrafo}
u_{\lambda_T^u \lambda_R^u}=U_{\lambda_T^u \lambda_R^u}+U_{-\lambda_T^u -\lambda_R^u}\,,\qquad v_{\lambda_T^u \lambda_R^u}=i(U_{\lambda_T^u \lambda_R^u}-U_{-\lambda_T^u -\lambda_R^u})\,.
\end{equation}
Thus, with the help of the function
\begin{eqnarray}
\hspace{-0.85cm}h_{\lambda_T^u \lambda_R^u,\lambda_T^{u'} \lambda_R^{u'},\lambda_T^{u''} \lambda_R^{u''}}(u,v)&:=&u_{\lambda_T^u \lambda_R^u}u_{\lambda_T^{u'} \lambda_R^{u'}}u_{\lambda_T^{u''} \lambda_R^{u''}}-u_{\lambda_T^u \lambda_R^u}v_{\lambda_T^{u'} \lambda_R^{u'}}v_{\lambda_T^{u''} \lambda_R^{u''}}\nonumber\\
 & &+v_{\lambda_T^u \lambda_R^u}u_{\lambda_T^{u'} \lambda_R^{u'}}v_{\lambda_T^{u''} \lambda_R^{u''}}+v_{\lambda_T^u \lambda_R^u}v_{\lambda_T^{u'} \lambda_R^{u'}}u_{\lambda_T^{u''} \lambda_R^{u''}}
\end{eqnarray}
the complex order parameter equations (\ref{ebopgkomplex}) are transformed to real ones as follows:
\begin{eqnarray}
&&\hspace{-1.2cm}\dot u_{\lambda_T^u \lambda_R^u}=u_{\lambda_T^u \lambda_R^u}\left[c_1\left(u_{\lambda_T^u \lambda_R^u}^2+v_{\lambda_T^u \lambda_R^u}^2\right)+c_2\left(u_{\lambda_T^u -\lambda_R^u}^2+v_{\lambda_T^u -\lambda_R^u}^2\right)\right.\nonumber\\
&&\hspace{-0.7cm}\left.+c_3\left(u_{\bar\lambda_T^u \lambda_R^u}^2+v_{\bar\lambda_T^u \lambda_R^u}^2+u_{\bar\lambda_T^u -\lambda_R^u}^2+v_{\bar\lambda_T^u -\lambda_R^u}^2\right)+c_5\left(u_{\lambda_T^u \bar\lambda_R^u}^2+v_{\lambda_T^u \bar\lambda_R^u}^2+u_{\lambda_T^u -\bar\lambda_R^u}^2+v_{\lambda_T^u -\bar\lambda_R^u}^2\right)\right.\nonumber\\
&&\hspace{-0.7cm}+c_7\left.\left(u_{\bar\lambda_T^u \bar\lambda_R^u}^2+v_{\bar\lambda_T^u \bar\lambda_R^u}^2+u_{\bar\lambda_T^u -\bar\lambda_R^u}^2+v_{\bar\lambda_T^u -\bar\lambda_R^u}^2\right)\right]+c_6 h_{\lambda_T^u -\lambda_R^u,\lambda_T^u \bar\lambda_R^u,\lambda_T^u -\bar\lambda_R^u}(u,v)\nonumber\\
&&\hspace{-0.7cm}+c_8\left[h_{\bar\lambda_T^u \bar\lambda_R^u,\bar\lambda_T^u \lambda_R^u,\lambda_T^u \bar\lambda_R^u}(u,v)+h_{\bar\lambda_T^u -\lambda_R^u,\lambda_T^u \bar\lambda_R^u,\bar\lambda_T^u -\bar\lambda_R^u}(u,v)\right.\nonumber\\
 & &\hspace{-0.7cm}\left.+h_{\bar\lambda_T^u -\bar\lambda_R^u,\lambda_T^u -\bar\lambda_R^u,\bar\lambda_T^u \lambda_R^u}(u,v)+h_{\bar\lambda_T^u -\lambda_R^u,\lambda_T^u -\bar\lambda_R^u,\bar\lambda_T^u \bar\lambda_R^u}(u,v)\right]\,,\label{ebreellu}
\end{eqnarray}
The equations for the amplitudes $v_{\lambda_T^u \lambda_R^u}$ have an identical structure as they are obtained from $(\ref{ebreellu})$ by exchanging the variables $u$ and $v$. 
It can be shown that the order parameter dynamics (\ref{ebreellu}) 
is governed by a potential \cite{thesis}. However, as the corresponding expression for the potential is lengthy, 
we will not discuss it here explicitly. 
Instead, we investigate different analytical cases which depend on the 
number of non-vanishing modes. In particular, we are interested in the emergence of retinotopical ordered projections between the planes. 

\subsection{Retinotopic Projections: Non-Vanishing Modes} \label{planerp}
We start with the assumption that only the amplitudes $u_j$, $v_j$ of one 
mode are different from zero. We consider the case $j=2$, but the other 
cases yield analogous results. The unstable part (\ref{UEXP}) reads 
\begin{equation}
U(t,r)=U_{10,-10}\exp\left[i 2\pi\left(\frac{t_1}{L_1^T}-
\frac{r_1}{L_1^R}\right)\right]+U_{-10,10}\exp\left[-i 2\pi\left(\frac{t_1}{L_1^T}
-\frac{r_1}{L_1^R}\right)\right]\,,
\end{equation}
which is equivalent to
\begin{equation}
U(t,r)=u_2 \cos\left[2\pi\left(\frac{t_1}{L_1^T}-\frac{r_1}{L_1^R}\right)\right]
+v_2 \sin\left[2\pi\left(\frac{t_1}{L_1^T}-\frac{r_1}{L_1^R}\right)\right]\,.
\end{equation}
The real order parameter equations (\ref{ebreellu}) reduce to
\begin{equation}
\dot u_2=\Lambda u_2+\frac{c_1}{4}u_2 (u_2^2+v_2^2)\,,\quad \dot v_2=\Lambda v_2+\frac{c_1}{4}v_2 (u_2^2+v_2^2)\,.
\end{equation}
We obtain constant phase-shift angles, which was already discussed in the case of strings (see Sec.~\ref{konstantephasenwinkel}). 
With $\xi=\sqrt{u_2^2+v_2^2}$ it follows
$\dot\xi=\Lambda\xi+(c_1/4)\xi^3$.
For the stationary case this leads to
$\xi=0 \,\mbox{ or }\, \xi=\sqrt{-4\Lambda/c_1}$.
We are only interested in the case $\xi\not=0$. This case corresponds to a 
retinotopy between $r_1$ and $t_1$, respectively. Thus, we have a retinotopic 
order only in one dimension and not in the whole plane.\\

Now we examine the question, if two modes are able to generate a retinotopic 
state in the plane. As a typical example we consider the case that $u_2$, $v_2$ as well 
as $u_8$, $v_8$ remain. Thus, the unstable part (\ref{UEXP}) has the complex decomposition 
\begin{eqnarray}
U(t,r)&=&U_{10,-10}\exp\left[i 2\pi\left(\frac{t_1}{L_1^T}-\frac{r_1}{L_1^R}
\right)\right]+U_{-10,10}\exp\left[-i 2\pi\left(\frac{t_1}{L_1^T}
-\frac{r_1}{L_1^R}\right)\right]\nonumber \\
&&+U_{01,0-1}\exp\left[i 2\pi\left(\frac{t_2}{L_2^T}-\frac{r_2}{L_2^R}
\right)\right]+U_{0-1,01}\exp\left[-i 2\pi\left(\frac{t_2}{L_2^T}
-\frac{r_2}{L_2^R}\right)\right]\,,
\label{eb2nichtverschw}
\end{eqnarray}
which due to (\ref{ebinvtrafo}) corresponds to the real decomposition
\begin{eqnarray}
U(t,r)&=&u_2 \cos\left[2\pi\left(\frac{t_1}{L_1^T}-\frac{r_1}{L_1^R}\right)\right]
+v_2 \sin\left[2\pi\left(\frac{t_1}{L_1^T}-\frac{r_1}{L_1^R}\right)\right]\nonumber\\
 & &+u_8 \cos\left[2\pi\left(\frac{t_2}{L_2^T}-\frac{r_2}{L_2^R}\right)\right]
+v_8 \sin\left[2\pi\left(\frac{t_2}{L_2^T}-\frac{r_2}{L_2^R}\right)\right]\,.
\end{eqnarray}
The real order parameter equations (\ref{ebreellu}) read
\begin{eqnarray}
\dot u_2&=&\left[\Lambda+\frac{c_1}{4}(u_2^2+v_2^2)+\frac{c_7}{4}(u_8^2+v_8^2)\right]u_2\,,\nonumber\\
\dot v_2&=&\left[\Lambda+\frac{c_1}{4}(u_2^2+v_2^2)+\frac{c_7}{4}(u_8^2+v_8^2)\right]v_2\,,\nonumber\\
\dot u_8&=&\left[\Lambda+\frac{c_1}{4}(u_8^2+v_8^2)+\frac{c_7}{4}(u_2^2+v_2^2)\right]u_8\,,\nonumber\\
\dot v_8&=&\left[\Lambda+\frac{c_1}{4}(u_8^2+v_8^2)+\frac{c_7}{4}(u_2^2+v_2^2)\right]v_8\,.
\end{eqnarray}
Again 
\begin{figure}[t!]
\centerline{\includegraphics[scale=0.9]{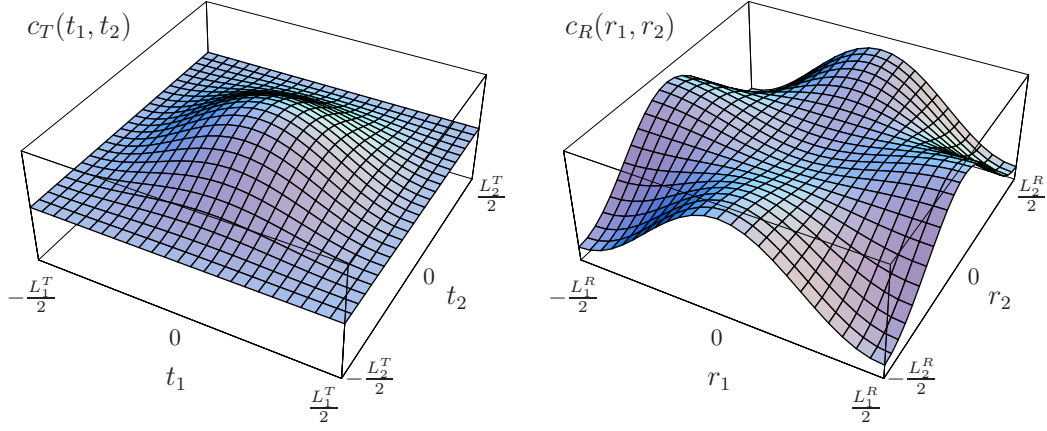}}
\caption{\label{koex1} \small Cooperativity functions for retina and 
tectum according to the special cases (\ref{eb9109}). 
The restriction to monotonically decreasing cooperativity functions has to be abandoned.}
\end{figure}
we obtain constant phase-shift angles.
With the amplitudes $\xi=\sqrt{u_2^2+v_2^2}\,,\eta=\sqrt{u_8^2+v_8^2}$
the following coupled equations result
\begin{equation} \label{ebxieta}
\dot\xi=\left(\Lambda+\frac{c_1}{4}\xi^2+\frac{c_7}{4}\eta^2\right)\xi\,,\qquad
\dot\eta=\left(\Lambda+\frac{c_1}{4}\eta^2+\frac{c_7}{4}\xi^2\right)\eta\,.
\end{equation}
We investigate under which conditions the two retinotopic modes coexist.
If $\xi,\eta\not=0$, we obtain from $\dot\xi=\dot\eta=0$ the relation
$(\xi^2-\eta^2)(c_1-c_7)=0$.
As we should minimize the restrictions for the coefficients $c_1$, $c_7$, we have in general $c_1\not= c_7$, 
so we conclude $\xi=\eta$. Inserting this result in (\ref{ebxieta}) 
for the stationary case leads to
\begin{equation} 
\label{ebxieta0}
\xi=\eta=\sqrt{-\frac{4\Lambda}{c_1+c_7}}\,.
\end{equation}
As the amplitudes $\xi,\eta$ have to be real and $\Lambda>0$, the coefficients $c_1, c_7$ have to fulfill the condition $c_1+c_7<0\,.$
Furthermore, we require that the coexistence of both modes is stable. 
To this end we consider the potential 
\begin{equation} 
\label{eb2modenpot}
V(\xi,\eta)=-\frac{\Lambda}{2}(\xi^2+\eta^2)-\frac{c_1}{16}(\xi^4+\eta^4)-\frac{c_7}{8}\xi^2 \eta^2\,,
\end{equation}
which reproduces
according to
\begin{equation}
\dot\xi=-\frac{\partial V(\xi,\eta)}{\partial \xi}\,,\quad \dot\eta=-\frac{\partial V(\xi,\eta)}{\partial \eta}
\end{equation}
the amplitude equations (\ref{ebxieta}). 
A stable state corresponds to a minimum of the potential (\ref{eb2modenpot}),
which leads to the condition $c_1 < c_7$.
With the explicit form of the coefficients $c_1$ and $c_7$ derived in Ref.~\cite{thesis} these considerations lead to the result, that either 
$\gamma^{20,20}$ or $\gamma^{11,11}$ could vanish.
If $\gamma^{20,20}=0$, the stability is guaranteed by
$-5\gamma^{11,11}/3<\gamma<-7\gamma^{11,11}$, whereas $\gamma^{11,11}=0$ demands 
$5\gamma^{20,20}/3<\gamma<-\gamma^{20,20}$.
As an example  
\begin{figure}[t!]
\centerline{\includegraphics[scale=0.9]{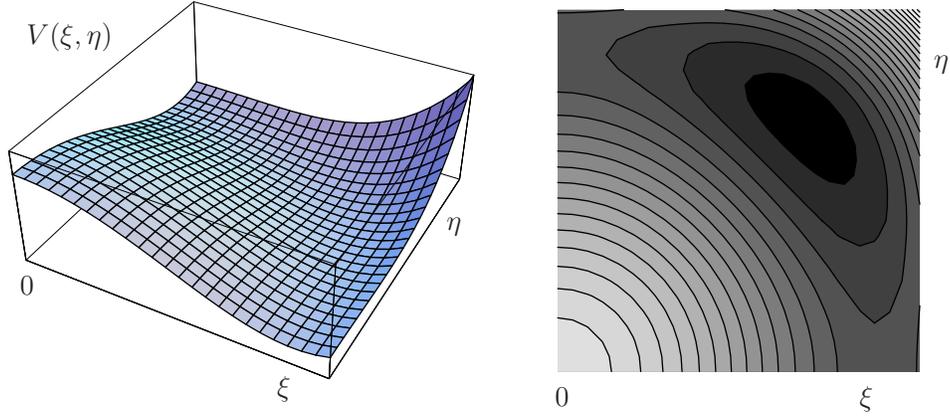}}
\caption{\label{potkoex} \small The potential (\ref{eb2modenpot}) 
for two coexistent retinotopic modes. According to (\ref{ebxieta0}) there is a maximum 
at $\xi=\eta=0$ and a minimum at $\xi=\eta$. The right plot shows the equipotential 
lines, which pronounces the extrema of $V$.}
\end{figure}
we consider the first case and assume a special form of the cooperativity functions (\ref{ebcoopexp}) with the expansion coefficients
\begin{equation} \label{eb9109}
f_{\pm 1 0}^T=f_{0 \pm 1}^T=f_{\pm 1 0}^R=f_{0 \pm 1}^R=0.1\,,\quad f_{\pm 1,\pm 1}^T=0.05\,,\quad f_{\pm 1,\pm 1}^R=-0.1\,,
\end{equation}
thus it follows $\gamma=0.01$ and $\gamma^{11,11}=-0.005$.  
However, although the cooperativity function of the tectum
is monotonically decreasing, the cooperativity function of the retina 
is not monotonic in this case, as is illustrated in Fig.~\ref{koex1}. The 
corresponding potential (\ref{eb2modenpot}) is shown in Fig.~\ref{potkoex}. 
\subsection{Center Manifold} \label{planecm}
In the next step we analyze the influence of the center manifold $S(U)$ 
for the special case (\ref{eb9109}). The
\begin{figure}[t!]
\centerline{\includegraphics[scale=0.9]{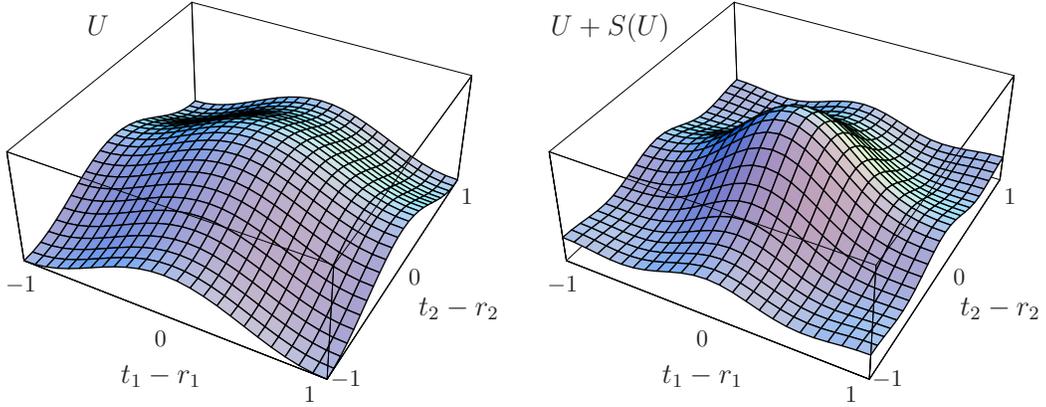}}
\caption{\label{huquadr} \small The contribution of the center manifold 
$S(U)$ leads to a more distinct concentration of the 
connection weight $w(t,r)\approx 1+U(t,r)+S(U(t,r))$ around $t=r$, as compared 
with the approximation $w(t,r)\approx 1+U(t,r)$.}
\end{figure}
connection weight can be represented as 
$w(t,r)=1+U(t,r)+S(U(t,r))$ (see Sec.~\ref{generalmodel}).
Using the relations (\ref{HN}) and (\ref{HNN}) 
the stable part is approximated in the second order.
We rewrite the result 
to real variables and eigenfunctions, where we set $L_1^{T,R}=L_2^{T,R}=2$. 
In this way we compare $U(t,r)$ with $U(t,r)+S(U(t,r))$
in Fig.~\ref{huquadr}.
It is evident that the retinotopic projection 
gets sharper due to the contribution of the center manifold, i.e.~the projection 
is maximal around the point $t=r$. This corresponds to the situation
found in the case of linear strings; the contribution of the higher 
modes have the tendency to support the emergence of retinotopic order. 
\section{Spheres} \label{spheres}
\begin{figure}[t!]
 \centerline{\includegraphics[scale=0.6]{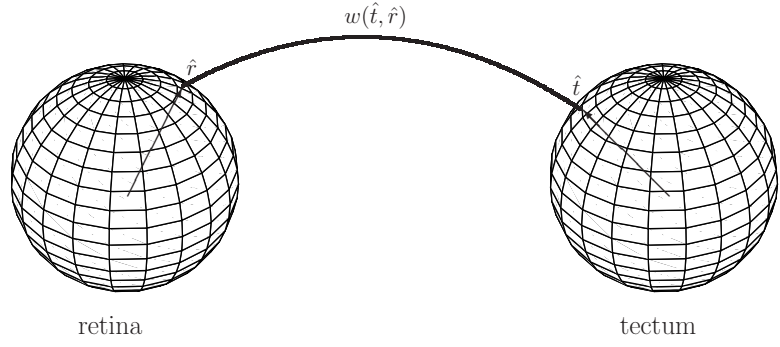}}
 \caption[Kugel]{\label{kugel} \small The cells of retina and tectum, which are assumed to be continuously distributed on unit spheres, are represented by their unit vectors $\hat r$ and $\hat t$, respectively. 
The two cell sheets are connected by positive connection weights $w(\hat t,\hat r)$.}
\end{figure}

In this Section we apply our general model \cite{gpw1} again to projections between two-dimensional
manifolds. Now, however, we consider manifolds with {\it constant positive curvature}. 
Typically, the retina represents approximately a hemisphere, whereas the tectum has an oval form \cite{goodhill,thesis}. 
Thus, it is biologically reasonable to model both cell sheets by spherical manifolds. 
Without loss of generality we assume that the two cell sheets for retina and tectum
are represented by the surfaces of two unit spheres, respectively. Thus, in our model, the corresponding continuously distributed cells are represented by unit vectors $\hat r$ and $\hat t$. Every ordered pair $(\hat t,\hat r)$ 
is connected by a positive connection weight $w(\hat t,\hat r)$ 
as is illustrated in Fig.~\ref{kugel}. 
The generalized H{\"a}ussler-von der Malsburg equations (\ref{HAUS}) for these connection weights are specified as follows
\begin{equation} 
\label{hslerkugel}
\dot w(\hat t,\hat r)=f(\hat t,\hat r,w)-\frac{w(\hat t,\hat r)}{8\pi} 
\hspace*{2mm} \int \! d\Omega_{t'}\,f(\hat t\,',\hat r,w)
-\frac{w(\hat t,\hat r)}{8\pi}\hspace*{2mm} 
\int \! d\Omega_{r'}\,f(\hat t,\hat r\,',w)\,,
\end{equation}
where the total growth rate is defined by
\begin{equation}
\label{GROsph}
f(\hat t,\hat r,w)=\alpha+w(\hat t,\hat r) \int \! d\Omega_{t'}
\int \! d\Omega_{r'} c_T(\hat t \cdot \hat t\,')\,
c_R(\hat r \cdot \hat r\,')\,w(\hat t\,',\hat r\,')\,.
\end{equation}
The integrations in (\ref{hslerkugel}) and (\ref{GROsph}) are performed over all
points $\hat t, \hat r$ on the spheres, where 
$d\Omega_t,d\Omega_r$ represent the differential solid angles of the
corresponding unit spheres. 
Note that the factors $8\pi$ in Eq.~(\ref{hslerkugel}) are twice the measure of the unit sphere.
After discussing the linear stability analysis around the homogeneous
solution of the generalized H{\"a}ussler-von der Malsburg equations (\ref{hslerkugel}) in
Subsec.~\ref{linanalys}, we perform the nonlinear synergetic analysis in Subsec.~\ref{nonlinanalys}, which yields the underlying order parameter equations
in the vicinity of the bifurcation. As in the case of Euclidean manifolds, 
we show that they have no quadratic terms, represent a potential dynamics,
and allow for retinotopic modes.
In Subsec.~\ref{11retino} we include the influence of higher modes upon the connection weights, which leads to
recursion relations for the corresponding amplitudes.
If we restrict ourselves to special cooperativity functions, 
the resulting recursion relations can be solved analytically by using again the method of 
generating functions, which was already applied in the case of strings. As a result of our analysis we obtain a perfect one-to-one
retinotopy if the global growth rate $\alpha$ is decreased to zero.

\subsection{Linear Analysis} \label{linanalys}
According to the general reasoning in Ref.~\cite{gpw1} we start with fixing the metric on the manifolds and determine the eigenfunctions
of the corresponding Laplace-Beltrami operator. Afterwards, we expand the cooperativity functions with respect to these eigenfunctions
and perform 
a linear analysis of the stationary uniform state.  

\subsubsection{Laplace-Beltrami Operator}
For the time being we neglect the distinction between retina and tectum,
because the following considerations are valid for both manifolds. 
Using spherical coordinates, we write
the unit vector on the sphere as $\hat x=(\sin\vartheta \cos\varphi,\sin\vartheta \sin\varphi,\cos\vartheta)$.
The Laplace-Beltrami operator 
for the sphere has the well-known form
\begin{equation} \label{einbettwinkel}
\Delta_{\vartheta,\varphi}=\frac{1}{\sin\vartheta}\,\frac{\partial}{\partial\vartheta}\left(\sin\vartheta \frac{\partial}{\partial\vartheta}\right)+\frac{1}{\sin^2\vartheta}\frac{\partial^2}{\partial\varphi^2}\,,
\end{equation}
whose eigenfunctions are
known to be given by spherical harmonics $Y_{lm}(\hat x)$, i.e.
\begin{equation}
\Delta_{\vartheta,\varphi}\,Y_{lm}(\hat x)=-l(l+1)Y_{lm}(\hat x)\,,
\end{equation}
which form a complete orthonormal system on the unit sphere.

\subsubsection{Cooperativity Functions}

The argument of the cooperativity functions $c(\hat x \cdot \hat x')$ is
the scalar product $\hat x\cdot \hat x'$ which takes values between $-1$ and $+1$. Therefore the cooperativity functions can be expanded in terms of
Legendre functions $P_l(\hat x\cdot \hat x')$, which form a complete orthogonal system on this interval \cite[7.221.1]{grad}: 
\begin{equation}
c(\hat x \cdot \hat x')=\sum_{l=0}^{\infty}\frac{2l+1}{4\pi}f_l\,P_l(\hat x \cdot \hat x')\,.\label{kuct}
\end{equation}
Using the Legendre addition theorem \cite{arf}
we arrive, for each manifold, at the expansion (\ref{CEX})
\begin{equation} \label{KU1}
c_T(\hat t \cdot \hat t')=\sum_{L=0}^{\infty}\sum_{M=-L}^L f_L^T Y_{LM}^T(\hat t\,) Y_{LM}^{T*}(\hat t'\,)\,,\quad c_R(\hat r \cdot \hat r')=\sum_{l=0}^{\infty}\sum_{m=-l}^l f_l^R Y_{lm}^R(\hat r) Y_{lm}^{R*}(\hat r'\,)\,.
\end{equation}
Note that the normalization of the cooperativity functions and the orthonormality relations
lead to the constraints $f_0^T=f_0^R=1$. 

\subsubsection{Eigenvalues}

According to Sec.~\ref{generalmodel}, a linear stability analysis around the stationary
uniform state leads to the eigenvalue problem of 
the linear operator (\ref{Loperator}). It has the eigenfunctions
$v_{L l}^{M m}(\hat t,\hat r)=Y_{LM}^T(\hat t\,) Y_{lm}^R(\hat r)$
and the spectrum of eigenvalues (\ref{ewsaite}).
By changing the uniform growth rate $\alpha$ in a suitable way, 
the real parts of some eigenvalues (\ref{ewsaite}) become positive
and the system can be driven to the neighborhood of an instability. 
Which eigenvalues (\ref{ewsaite}) become unstable
in general depends on the respective values
of the given expansion coefficients $f_L^T$, $f_l^R$. 
If we assume monotonically decreasing expansion coefficients $f_L^T$, $f_l^R$,
i.e.
\begin{equation}
1=f_0^T\geq f_1^T\geq f_2^T\geq \cdots \geq 0\,,\qquad 1=f_0^R\geq f_1^R\geq f_2^R\geq \cdots \geq 0\,,
\end{equation}
the maximum eigenvalue in (\ref{ewsaite}) is given by
$\Lambda_{\rm max}=\Lambda_{11}^{M m}=-\alpha+f_1^T f_1^R$.
Thus, the instability occurs when the global growth rate reaches
its critical value $\alpha_c=f_1^T f_1^R$.
At this instability point all nine modes with $(L^u,l^u)=(1,1)$ and $M^u=0,\pm 1$, $m^u=0,\pm 1$ become unstable, where we have introduced the index $u$ for the unstable modes. 

\subsection{Nonlinear Analysis} \label{nonlinanalys}
In this subsection we specialize the generic order parameter equations of Ref.~\cite{gpw1}
to unit spheres. We observe
that the quadratic term vanishes and derive selection rules for the appearance of cubic
terms. Furthermore, we essentially simplify the calculation of the order parameter equations by taking into account the symmetry properties of the cubic terms.
We show that the order parameter equations
represent a potential dynamics, and determine the underlying potential. 
\subsubsection{General Structure of Order Parameter Equations} \label{GSOPE}
The expansion of the unstable modes (\ref{UEXP}) reads
$U(\hat t,\hat r)=U_{11}^{M^u m^u} Y_{1 M^u}^T (\hat t\,) Y_{1 m^u}^R (\hat r\,)$
and, correspondingly, the contribution of the stable modes (\ref{SEXP}) is given by $S(\hat t,\hat r)=S_{L l}^{M m} Y_{L M}^T (\hat t\,) Y_{l m}^R (\hat r\,)\,.$
Note that the latter summation is performed over all parameters $(L,l)$ except for $(L^u,l^u)=(1,1)$, i.e. from now on the parameters $(L,l)$ stand for the stable modes alone. 
The resulting order parameter equations (\ref{OPE}) read
\begin{eqnarray}
\dot U^{M^u m^u} &=&\Lambda \,
U^{M^u m^u} +
A_{M^u, M^u{}' M^u{}''}^{m^u, m^u{}' m^u{}''}
\,U^{M^u{}' m^u{}'} \, U^{M^u{}'' m^u{}''} 
\nonumber\\
&&+ B_{M^u, M^u{}' M^u{}'' M^u{}'''}^{m^u{},
m^u{}' m^u{}'' m^u{}'''} U^{M^u{}' m^u{}'}
\, U^{M^u{}'' m^u{}''}\, U^{M^u{}''' m^u{}'''} \,.
\label{OPEsph}
\end{eqnarray}
With the integrals (\ref{abkurzI})
\begin{eqnarray} 
I_{l,l^{(1)} l^{(2)} \ldots  l^{(n)}}^{m,m^{(1)} m^{(2)} \dots m^{(n)}}
&=&\hspace*{1mm}\int\! d\Omega_x\,
Y_{lm}^*(\hat x)\,Y_{l^{(1)}\,m^{(1)}}(\hat x)\,Y_{l^{(2)}\,m^{(2)}}(\hat x)\,\cdots\,
Y_{l^{(n)}\,m^{(n)}}(\hat x)\,,\label{abkurzIsph}\\
J_{l^{(1)} l^{(2)} \ldots  l^{(n)}}^{m^{(1)} m^{(2)} \ldots m^{(n)}}
&=&\hspace*{1mm}\int\!d\Omega_x\,
Y_{l^{(1)}\,m^{(1)}}(\hat x)\, Y_{l^{(2)}\,m^{(2)}}(\hat x)\,\cdots\,
Y_{l^{(n)}\,m^{(n)}} (\hat x) \label{abkurzJsph}
\end{eqnarray}
the quadratic coefficients (\ref{AA}) read
$A_{M^u, M^u{}' M^u{}''}^{m^u, m^u{}' m^u{}''}  = f_1^T \,
f_1^R \, I_{1,1\,1}^{M^u,M^u{}' M^u{}''}
\, I_{1,1\,1}^{m^u,m^u{}' m^u{}''} \,,$
whereas the cubic coefficients (\ref{BB}) are
\begin{eqnarray}
&&\hspace{-1.6cm}B_{M^u, M^u{}' M^u{}'' M^u{}'''}^{m^u{}, m^u{}' m^u{}'' m^u{}'''}
= - \frac{1}{8\pi}\,f_1^T \, f_1^R
\left(I_{1,1\,1\,1}^{M^u,M^u{}' M^u{}'' M^u{}'''}
\,\delta_{m^u m^u{}'} \,J_{1\,1}^{m^u{}'' m^u{}'''}
+ I_{1,1\,1\,1}^{m^u,m^u{}' m^u{}'' m^u{}'''} \right.\nonumber\\
&&\hspace{-1.5cm}\left.\times\delta_{M^u M^u{}'} \,
J_{1\,1}^{M^u{}'' M^u{}'''} \right)
+\left\{ \left[ f_L^T \, f_l^R +f_1^T \, f_1^R \, \right]
I_{1,1\,L}^{M^u,M^u{}' M} \,
I_{1,1\,l}^{m^u,m^u{}' m}
- \frac{1}{4\sqrt{\pi}} \,
\left[\delta_{L0}\,\delta_{M0} \delta_{M^u M^u{}'}
\right.\right.\nonumber\\
&&\hspace{-1.5cm}\left.\left.\times\left( 1+ f_l^R \right)
I_{1,1\,l}^{m^u,m^u{}' m}
\right.\right. \left. \left.
+\delta_{l0}\,\delta_{m0}\delta_{m^u m^u{}'}
\, \left( 1+f_L^T\right)
\, I_{1,1\,L}^{M^u,M^u{}' M}
\right] \right\} H_{Ll}^{Mm,M^u{}'' m^u{}'' M^u{}''' m^u{}'''} \,
\,.
\label{BBsph}
\end{eqnarray}
The center manifold coefficients 
$H_{Ll}^{Mm,M^u m^u M^u{}' m^u{}'}$
read 
\begin{eqnarray}
\label{HNNsph}
H_{Ll}^{Mm,M^u m^u M^u{}' m^u{}'}  &=& 
\frac{f_1^T f_1^R}{2\Lambda-\Lambda_{Ll} }
\Bigg[ I_{L,1\,1}^{M,M^u M^u{}'} \, 
I_{l,1\,1}^{m,m^u m^u{}'} 
- \frac{1}{4\sqrt{\pi}} \, \left(J_{1\,1}^{M^u M^u{}'} 
\,I_{l,1\,1}^{m,m^u{}' m^u{}''} \,\delta_{L0} \right.\nonumber\\  
 & &\left.+J_{1\,1}^{m^u m^u{}'} \,I_{L,1\,1}^{M,M^u M^u{}'} 
\,\delta_{l0} \right) \Bigg] \,. 
\label{QRESsph}
\end{eqnarray}

\subsubsection{Integrals}
The order parameter equations contain the following integrals: $J_{11}^{m' m''},I_{l,11}^{m,m' m''},I_{1,1 l}^{m,m' m''}$, $I_{1,111}^{m,m' m'' m'''}$.
For the first integral one easily obtains
$J_{1 1}^{m' m''}=(-1)^{m'}\delta_{m',-m''}$. Integrals over three and four spherical harmonics can be calculated using the Wigner-Eckart-theorem,
\begin{equation} \label{3fachint}
I_{l,l' l''}^{m,m' m''}=\sqrt{\frac{(2l'+1)(2l''+1)}{4\pi(2l+1)}}\,C(l',0,l'',0|l,0)\,C(l',m',l'',m''|l,m)\,,
\end{equation}
whence for $l'=l''=1$ it follows
\begin{equation} \label{1046}
I_{l,1 1}^{m,m' m''}=\frac{3}{\sqrt{4\pi(2l+1)}}\,C(1,0,1,0|l,0)\,C(1,m',1,m''|l,m)\,.
\end{equation}
As the Clebsch-Gordan coefficients $C(l_1,0,l_2,0|l_3,0)$ vanish if the sum
$l_1+l_2+l_3$ is odd \cite{heine},
we obtain $I_{1,11}^{m,m' m''}=0$.
Thus, the quadratic contribution to the
order parameter equations (\ref{OPEsph}) vanishes, by analogy with Euclidean manifolds.
Non-vanishing integrals (\ref{1046}) can only occur for $l=0$ and $l=2$. 
Furthermore, the integrals $I_{1,1 l}^{m,m' m''}$ follow from
$I_{1,1 l}^{m,m' m''}=(-1)^{m'+m''}I_{l,1 1}^{-m'',-m\,m'}$.
Integrals over four spherical harmonics can also be calculated using the Wigner-Eckart-theorem, and the result is
\begin{eqnarray} \label{sphere38}
I_{l,l' l'' l'''}^{m,m' m'' m'''}&=&\sum_{l_3=|l''-l'''|}^{l''+l'''}\sum_{m_3=-l_3}^{l_3}\sqrt{\frac{(2l''+1)(2l'''+1)}{4\pi(2l_3+1)}}\,C(l'',0,l''',0|l_3,0)\nonumber\\
&&\times C(l'',m'',l''',m'''|l_3,m_3) I_{l,l' l_3}^{m,m' m_3}\,.
\end{eqnarray} 
Specialyzing (\ref{sphere38}) to $l=l'=l''=l'''=1$ and taking into account (\ref{3fachint}) leads to
$I_{1,1\,1\,1}^{m,m' m'' m'''}\propto\delta_{m'+m''+m''',m}$.
Thus, we obtain the selection rule that the nonvanishing integrals $I_{1,1\,1\,1}^{m,m' m'' m'''}$ fulfill the condition $m'+m''+m'''=m$. 

\subsubsection{Order Parameter Equations} 
To simplify the calculation of the cubic coefficients (\ref{BBsph}) in the order parameter equations (\ref{OPEsph}), we perform some 
basic considerations which lead to helpful symmetry properties.
To this end we start with replacing $m^u$ by $-m^u$ which yields 
$I_{1,1\,1\,1}^{m^u,m^{u'} m^{u''} m^{u'''}}=I_{1,1\,1\,1}^{-m^u,-m^{u'}-m^{u''}-m^{u'''}}$. Corresponding symmetry relations can also be derived for the other
terms in (\ref{BBsph}). Therefore, we conclude that the order parameter equation for $U^{-M^u -m^u}$ is
obtained from that of $U^{M^u m^u}$ by negating all 
indices $M^u$ and $m^u$ with unchanged factors. 
Thus, instead of explicitly calculating nine order parameter equations, 
it is sufficient to restrict oneself determining the order parameter equations for $U^{00}$, $U^{10}$, $U^{01}$, and $U^{11}$. The remaining five order parameter equations follow instantaneously from those
by applying the symmetry relations, as is further worked out in Ref.~\cite{thesis}.\\

To investigate how the complex order parameter equations contribute to the
one-to-one retinotopy, we transform them to real variables according to
\begin{eqnarray}
u_0=U^{00}/\sqrt{2}\,,&u_1=(U^{11}+U^{-1-1})/2\,,&u_2=i(U^{11}-U^{-1-1})/2\,,\nonumber\\
u_3=(U^{1-1}+U^{-11})/2\,,&u_4=i(U^{1-1}-U^{-11})/2\,,&u_5=(U^{01}-U^{0-1})/2\,,\nonumber\\
u_6=i(U^{01}+U^{0-1})/2\,,&u_7=(U^{10}-U^{-10})/2\,,&u_8=i(U^{10}+U^{-10})/2\,. 
\end{eqnarray}
The resulting real order parameter equations turns out to follow according to
\begin{equation} \label{kupotdynamik}
\dot u_i=-\frac{\partial V(\{u_j\})}{\partial u_i}
\end{equation}
from the potential
\begin{eqnarray}
&&\hspace{-1cm}V(\{u_j\})=-\frac{\Lambda}{2}\sum_{j=0}^8u_j^2-\frac{\beta_1}{2}u_0^4-\bar\beta_2 u_0^2(u_5^2+u_6^2)-\beta_2 u_0^2(u_7^2+u_8^2)-\beta_3 u_0^2(u_1^2+u_2^2+u_3^2+u_4^2)\nonumber\\
 & &\hspace{-1cm}-\sqrt{2}\beta_4u_0(u_1u_5u_7+u_2u_5u_8+u_2u_6u_7+u_4u_6u_7-u_1u_6u_8-u_3u_5u_7-u_4u_5u_8-u_3u_6u_8)\nonumber\\
 & &\hspace{-1cm}-\beta_5(u_5^2-u_6^2)(u_1u_3+u_2u_4)-\beta_5(u_7^2-u_8^2)(u_1u_3-u_2u_4)-2\beta_5u_7u_8(u_1u_4+u_2u_3)\nonumber\\
 & &\hspace{-1cm}-2\beta_5u_5u_6(u_2u_3-u_1u_4)+\frac{1}{2}[\beta_6(u_5^2+u_6^2)+\bar\beta_6(u_7^2+u_8^2)](u_1^2+u_2^2+u_3^2+u_4^2)\nonumber\\
 &&\hspace{-1cm}-\frac{\beta_7}{2}(u_1^2+u_2^2)(u_3^2+u_4^2)-\beta_3 (u_5^2+u_6^2)(u_7^2+u_8^2)\nonumber\\
&&\hspace{-1cm}-\frac{\beta_8}{4}\left[(u_1^2+u_2^2)^2+(u_3^2+u_4^2)^2\right]+\frac{\beta_9}{4}(u_5^2+u_6^2)^2+\frac{\bar\beta_9}{4}(u_7^2+u_8^2)^2\,. \label{kupot}
\end{eqnarray}
The dependence of the coefficients $\beta_i$ and $\bar\beta_i$ on the expansion coefficients $f_l$ and the control parameter $\alpha$
is found in Ref.~\cite{thesis}.
Naturally, a complete analytical determination of all stationary states of the real order parameter equations (\ref{kupotdynamik}) is
impossible. However, we are able to demonstrate that certain stationary states
admit for retinotopic modes. 
\subsubsection{Special Case}

To this end we consider the special case 
where $u_1,u_2,u_5,u_6,u_7,u_8$ vanish. Then the order parameter equations (\ref{kupotdynamik}) with (\ref{kupot})
for the non-vanishing amplitudes $u_0$, $u_3$, $u_4$ reduce to 
\begin{eqnarray}
\dot u_0&=&\Lambda u_0+2 \beta_1 u_0^3+2\beta_3 (u_3^2+u_4^2)u_0\,,\nonumber\\
\dot u_3&=&\Lambda u_3+2\beta_3 u_0^2 u_3+\beta_8 (u_3^2+u_4^2)u_3\,,\nonumber\\
\dot u_4&=&\Lambda u_4+2\beta_3 u_0^2 u_4+\beta_8 (u_3^2+u_4^2)u_4\,.
\end{eqnarray}
Due to the relation $\dot u_3/u_3=\dot u_4/u_4$
one obtains constant phase-shift angles, i.e. it holds $u_3\propto u_4$. 
Therefore, the system of three coupled differential equations can be reduced to two variables. 
To this end we introduce the new variable 
\begin{equation} \label{kuneuvariab} 
\xi=\sqrt{u_3^2+u_4^2}\,,
\end{equation}
which leads to
\begin{equation} \label{kuu0glg}
\dot u_0=\Lambda u_0+2 \beta_1 u_0^3+2\beta_3 \xi^2 u_0\,,\quad
\dot \xi=\Lambda \xi+2\beta_3 u_0^2 \xi +\beta_8 \xi^3\,.
\end{equation}
The stationary solution, which corresponds to a coexistence of the two modes, 
is given by
\begin{equation} \label{kuu0xilsg}
u_0^2=-\frac{\Lambda}{2(\beta_3+\beta_8)}\,,\quad\xi^2=-\frac{\Lambda}{\beta_3+\beta_8}\,,
\end{equation}
where we used the relation $\beta_8=\beta_1+\beta_3$ \cite{thesis}. Demanding real amplitudes $u_0$, $\xi$ leads to the coexistence condition
$\beta_3+\beta_8<0$.
Furthermore, we require stability for this state. Therefore we consider
the corresponding potential $V(u_0,\xi)$, which can be read off from (\ref{kupot}) and (\ref{kuneuvariab}):
\begin{equation}
V(u_0,\xi)=-\frac{\Lambda}{2}(u_0^2+\xi^2)-\frac{\beta_1}{2}u_0^4-\beta_3 u_0^2\xi^2-\frac{\beta_4}{4}\xi^4\,.
\end{equation}
Stable states correspond to a minimum of $V$, which leads to the conditions 
$2\beta_3-\beta_8>0\,,\quad \beta_3-\beta_8>0$.
If all three inequalities are valid, both the $u_0$- and the $\xi$-mode coexist.  
If we set $u_4=0$, without loss of generality, the solution reads in complex variables 
according to (\ref{kuneuvariab}) 
\begin{equation}
U^{00}=\sqrt{-\frac{\Lambda}{\beta_3+\beta_8}}\,,\qquad U^{1-1}=U^{-11}=-\sqrt{-\frac{\Lambda}{\beta_3+\beta_8}}\,.
\end{equation}
Thus, the unstable part, specified in Sec.~\ref{GSOPE}, is given by
\begin{equation} \label{sphere66}
U(\hat t,\hat r)=\sqrt{-\frac{\Lambda}{\beta_3+\beta_8}}\,\Big[Y_{10}^T(\hat t\,) Y_{10}^R(\hat r)-Y_{11}^T(\hat t\,)Y_{1-1}^R(\hat r)-Y_{1-1}^T(\hat t\,)Y_{11}^R(\hat r)\Big]\,.
\end{equation}
Using the Legendre addition theorem reduces (\ref{sphere66}) to
\begin{equation} \label{kupl1}
U(\hat t,\hat r)=\sqrt{-\frac{\Lambda}{\beta_3+\beta_8}}\,P_1(\hat t\cdot\hat r)
\end{equation}
with $P_1(\hat t\cdot \hat r)=\hat t \cdot \hat r$. Thus, the unstable part is minimal, if $\hat t$ and $\hat r$ are antiparallel,
i.e. the distance of the corresponding points on the unit sphere is maximum.
Decreasing the angle between $\hat t$ and $\hat r$ leads to increasing 
values of $U(\hat t,\hat r)$, and the maximum occurs for parallel unit vectors.
This justifies calling the mode (\ref{kupl1}) retinotopic.
\subsection{One-to-One Retinotopy} \label{11retino}
Now we investigate whether the generalized H{\"a}ussler-von der Malsburg equations (\ref{hslerkugel})
describe the emergence of a perfect one-to-one retinotopy between two spheres. 
To this end we follow the unpublished suggestions of Ref.~\cite{Malsburg4} and 
treat systematically the contribution
of higher modes. Because the Legendre functions form a complete orthogonal system for functions defined on the interval $[-1,+1]$, 
their products can always be written as linear combinations 
of Legendre functions. This motivates that the influence
of higher modes upon the connection weights, which obey
the generalized H{\"a}ussler-von der Malsburg equations (\ref{hslerkugel}), can be 
included by the ansatz
\begin{equation} \label{wagneransatz}
w(\sigma)=\sum_{l=0}^{\infty}(2l+1)Z_l P_l(\sigma)\,,
\end{equation}
where the amplitudes $Z_l$ are time dependent. 

\subsubsection{Recursion Relations}
Inserting (\ref{wagneransatz}) into the generalized H{\"a}ussler-von der Malsburg equations (\ref{hslerkugel}) and performing the integrals 
over the respective unit spheres leads to
\begin{eqnarray} \label{wagnerstern}
\sum_{l=0}^{\infty}(2l+1)\dot Z_l P_l(\sigma)&=&\alpha\left[1-\sum_{l=0}^{\infty}(2l+1)Z_l P_l(\sigma)\right]\nonumber\\
 & &+\sum_{l=0}^{\infty}(2l+1)Z_l P_l(\sigma)\sum_{l'=0}^{\infty}(2l'+1)Z_{l'}f_{l'}^T f_{l'}^R [P_{l'}(\sigma)-Z_{l'}]\,.
\end{eqnarray}
The products of Legendre functions occuring in (\ref{wagnerstern}) can be 
reduced to linear combinations of single Legendre functions according to 
the standard decomposition \cite[8.915]{grad}
\begin{equation}
P_l(\sigma)P_{l'}(\sigma)=\sum_{k=0}^{l}A_{l,l',k}P_{l+l'-2k}(\sigma)\,,\quad l\leq l'
\end{equation}
with the coefficients
\begin{equation} \label{wagnerA}
A_{l,l',k}=
\frac{(2l'+2l-4k+1)\,a_{l'-k}a_k a_{l-k}}
{(2l'+2l-2k+1)\,a_{l+l'-k}}\,,
\quad a_k=\frac{(2k-1)!!}{k!}\,.
\end{equation}
Thus, contributions to the polynomial $P_{\tilde l}(\sigma)$ only occur iff the relation $k=(l+l'-\tilde l)/2$
is fulfilled. Furthermore, using the orthonormality relation of the polynomials yields the following recursion relation for the amplitudes $Z_l$:
\begin{eqnarray}
(2l+1)\dot Z_l&=&\alpha[\delta_{l,0}-(2l+1)Z_l]-(2l+1)Z_l(Z_0^2+3f_1^T f_1^R Z_1^2)\nonumber\\
 & &+\sum_{l'=0}^{\infty}(2l'+1)Z_{l'}\left[\sum_{l''=0}^l (2l''+1)Z_{l''}f_{l''}^T f_{l''}^R \sum_{k=0}^{l''}A_{l',l'',k}\delta_{k,(l'+l''-l)/2}\right.\nonumber\\
 & &\left.+\sum_{l''=l'+1}^{\infty}(2l''+1)Z_{l''}f_{l''}^T f_{l''}^R \sum_{k=0}^{l'}A_{l',l'',k}\delta_{k,(l'+l''-l)/2} \right]\,. \label{128}
\end{eqnarray}
Note that Eq.~(\ref{128}) cannot be solved analytically for arbitrary expansion coefficients $f_l^T$, $f_l^R$ 
of the cooperativity functions. Therefore, we restrict ourselves from now on to
a special case.

\subsubsection{Special Cooperativity Functions}
For simplicity we assume for the 
cooperativity functions (\ref{kuct}) that
$f_0=1\,,f_1\not=0\,,f_l=0\, \mbox{ for } l\not=0,\pm 1$.
With this choice the recursion relation (\ref{128}) for $l=0$ reduces to
\begin{equation} \label{wagnerz0}
\dot Z_0=-(\alpha+Z_0^2+3\gamma Z_1^2)(Z_0-1)\,,
\end{equation}
where we have used again the abbreviation $\gamma=f_1^T f_1^R$.
For $l\not=0$, by taking into account (\ref{wagnerA}), we obtain 
\begin{equation} \label{wagner3stern}
\dot Z_l=-(\alpha+Z_0^2+3\gamma Z_1^2)Z_l+Z_0 Z_l+3\gamma Z_1\,\frac{lZ_{l-1}+(l+1)Z_{l+1}}{2l+1}\,.
\end{equation}
The long-time behavior of the system corresponds to its stationary states. 
They are determined by $Z_0=1$ from (\ref{wagnerz0}), whereas
(\ref{wagner3stern}) leads to a nonlinear recursion relation for the amplitudes $Z_l$ with $l\not=0$. However, by introducing the variable
\begin{equation} \label{wagnerualpha}
u=\frac{\alpha+3\gamma Z_1(u)^2}{3\gamma Z_1(u)}\,,
\end{equation}
this nonlinear recursion relation can be formally transformed into the linear one
\begin{equation} \label{wagnerrek}
(l+1)Z_{l+1}(u)=(2l+1)uZ_l(u)+lZ_{l-1}(u)\,,\quad l\geq 1\,.
\end{equation}
Thus, determining the stationary solution of the nonlinear recursion relation (\ref{wagner3stern}) amounts to
solving the linear recursion relation (\ref{wagnerrek}) for $Z_l(u)$ in such
a way that the self-consistency condition (\ref{wagnerualpha}) is fulfilled.

\subsubsection{Generating Function}
To determine the amplitudes $Z_l(u)$ we calculate their
generating function
\begin{equation} \label{wagnererz}
E(x,u)=\sum_{l=0}^{\infty}Z_l(u)x^l\,,
\end{equation}
where we have the normalization 
\begin{equation} \label{wagnerex0}
E(0,u)=Z_0(u)=1\,.
\end{equation}
Multiplying both sides of (\ref{wagnerrek}) with $x^l$ and summing over $l\geq 1$
leads to an inhomogeneous nonlinear partial differential equation of first order for the generating function: 
\begin{equation} \label{wagnerinh}
(x^2-2ux+1)\,\frac{\partial E(x,u)}{\partial x}=(u-x)E(x,u)+Z_1(u)-u\,.
\end{equation}
Using the normalization condition (\ref{wagnerex0}) its solution is given by
\begin{equation} \label{wagnererzeugende}
E(x,u)=\frac{1+[Z_1(u)-u]\ln\displaystyle{\frac{\sqrt{x^2-2ux+1}+x-u}{1-u}}}{\sqrt{x^2-2ux+1}}\,.
\end{equation}

\subsubsection{Decomposition}
We now determine the unknown amplitudes $Z_l(u)$.
From the mathematical literature
it is well-known that the recursion relation (\ref{wagnerrek}) holds both for the
Legendre functions of first kind $P_l(u)$ and second kind $Q_l(u)$, respectively \cite{grad}. 
Thus, we expect that the generating function (\ref{wagnererzeugende}) can be
represented as a linear combination of the generating functions of the Legendre
functions of both first and second kind, 
which are given by \cite[8.921]{grad} and \cite[8.791.2]{grad}:
\begin{eqnarray}
E_P(x,u)&=&\sum_{l=0}^{\infty}P_l(u)x^l=\frac{1}{\sqrt{x^2-2ux+1}}\,,\label{Pgen}\\
E_Q(x,u)&=&\sum_{l=0}^{\infty}Q_l(u)x^l=\frac{\ln\displaystyle{\frac{\sqrt{x^2-2ux+1}+u-x}{\sqrt{u^2-1}}}}{\sqrt{x^2-2ux+1}}\,.\label{Qgen}
\end{eqnarray}
Indeed, taking into account the explicit form of the Legendre function of second kind for $l=0$~\cite{arf}
\begin{equation} \label{wagnerq0}
Q_0(u)=\frac{1}{2}\ln \frac{u+1}{u-1}\,,
\end{equation}
the generating function (\ref{wagnererzeugende}) decomposes according to
\begin{equation}
E(x,u)=\{1+[Z_1(u)-u]Q_0(u)\}E_P(x,u)-[Z_1(u)-u]E_Q(x,u)\,.
\end{equation}
Inserting (\ref{Pgen}), (\ref{Qgen}) and performing a comparison with (\ref{wagnererz}) then yields the result
\begin{equation} \label{wagnerzl}
Z_l(u)=\{1+[Z_1(u)-u]Q_0(u)\}P_l(u)-[Z_1(u)-u]Q_l(u)\,.
\end{equation}
Thus, the amplitudes $Z_l(u)$ turn out to be linear combinations of $P_l(u)$ and $Q_l(u)$. 
To fix the yet undetermined amplitude $Z_1(u)$ in the expansion coefficients of
(\ref{wagnerzl}), we have to take into account the boundary condition that
the sum in the ansatz (\ref{wagneransatz}) has to converge.

\subsubsection{Boundary Condition}
\begin{figure}[t!]
 \centerline{\includegraphics[scale=0.85]{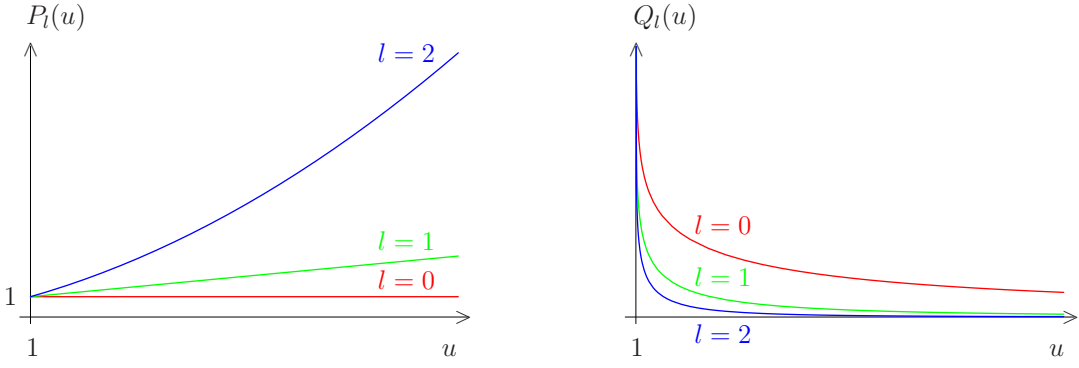}}
 \caption[Legendre-Polynome]{\label{pqlegendre} \small The Legendre functions of first and second kind $P_l(u)$ and $Q_l(u)$ for $u>1$. 
We have $P_l(1)=1$, whereas $Q_l(u)$ diverges for $u\downarrow 1$. 
Important for the boundary condition of $Z_l(u)$ is the different behavior
for increasing values of $l$: $P_l(u)$ diverges, whereas $Q_l(u)$ converges to zero.}
\end{figure}
Because the Legendre functions $P_l(\sigma)$ do not vanish with increasing $l$, 
we must require that $Z_l(u)$ vanishes in the limit $l\to \infty$.
The series of Legendre functions of first kind $P_l(u)$ with fixed $u>1$ diverges for $l\to \infty$ according to \cite[8.917]{grad}:
$P_0(u)<P_1(u)<P_2(u)<\ldots<P_n(u)<\ldots\,,\, u>1$.
The Legendre functions of second kind $Q_l(u)$, however, converge to zero (see Fig.~\ref{pqlegendre}). Thus, performing the limit
$l\to \infty$ in Eq.~(\ref{wagnerzl})
and using the explicit form \cite{arf} $Q_1(u)=uQ_0(u)-1$
it follows that $Z_1(u)$ is fixed according to
$Z_1(u)=Q_1(u)/Q_0(u)$.
With this we obtain that 
the result (\ref{wagnerzl}) finally reads
\begin{equation} \label{wagnerzlq}
Z_l(u)=\frac{Q_l(u)}{Q_0(u)}\,,
\end{equation}
which is not valid only for $l\not=0$ but also for $l=0$ due to (\ref{wagnerex0}).

\subsubsection{Connection Weight}
Inserting (\ref{wagnerzlq}) into (\ref{wagneransatz}) yields the following 
solution for the connection weight:
\begin{equation}
w(\sigma)=\frac{1}{Q_0(u)}\sum_{l=0}^{\infty}(2l+1)Q_l(u)P_l(\sigma)\,.
\end{equation}
Using the identity \cite[8.791.1]{grad} 
and (\ref{wagnerq0}), we obtain for the connection weight 
\begin{equation} \label{117}
w(\sigma)=\frac{2}{u-\sigma}\left(\ln\frac{u+1}{u-1}\right)^{-1}\,.
\end{equation}
Note that integrating (\ref{117}) over the unit sphere leads to $4\pi$, 
i.e. the total connection weight coincides with the measure $M$. \\

On the other hand we have to take into account that the self-consistency 
condition (\ref{wagnerualpha}) yields an explicit relation between the variable $u$ and the control parameter $\alpha$. 
Indeed, we infer from (\ref{wagnerualpha}) 
and (\ref{wagnerzlq}) the following transcendental relation between
$\alpha$ and $u$
\begin{equation} \label{wagnera2g}
\frac{\alpha}{\gamma}=-\frac{2}{3}\left(\ln\frac{u+1}{u-1}\right)^{-1}\left[2\left(\ln\frac{u+1}{u-1}\right)^{-1}-u\right]\,, 
\end{equation}
which is depicted in Fig.~\ref{alphavonu}a. 

\subsubsection{Limiting Cases}
\begin{figure}[t!]
\includegraphics[width=0.45\textwidth]{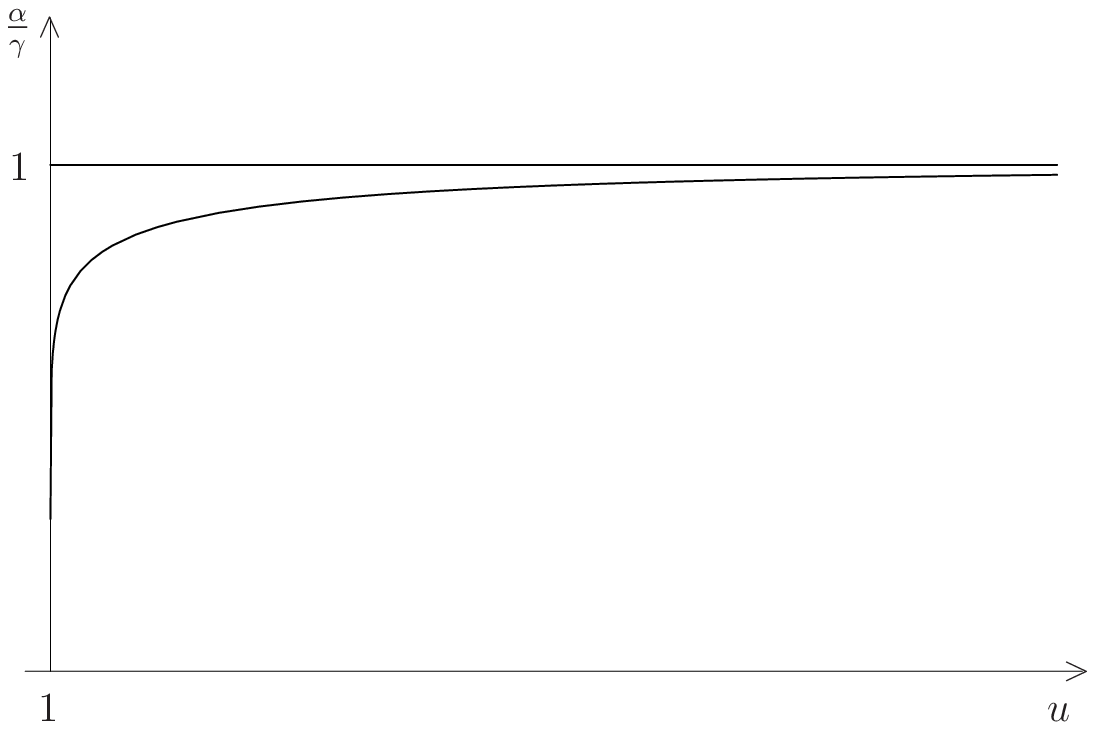}~a)
\hfil
\includegraphics[width=0.45\textwidth]{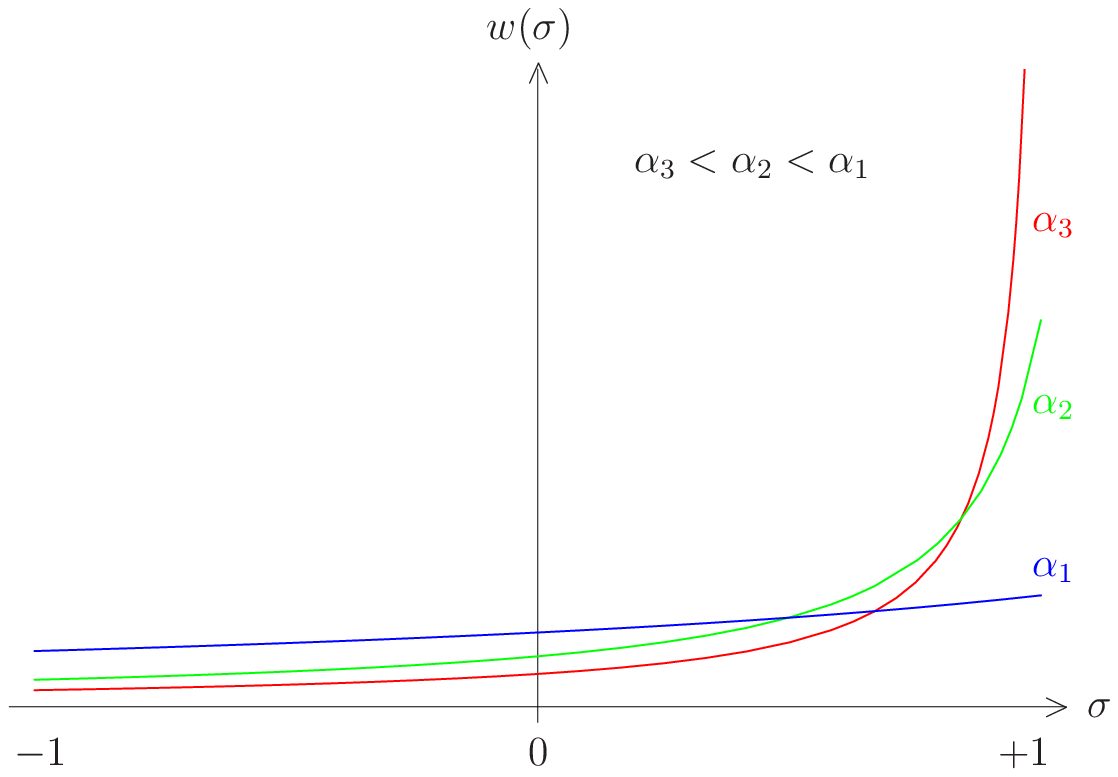}~b)
\caption[Kontrollparameter]{\label{alphavonu} \small a) Relation (\ref{wagnera2g}) between the control parameter $\alpha$ and the variable $u$. b) The 
connection weight for different values of the control parameter $\alpha$. For decreasing values of $\alpha$ the connection weight around $\sigma=+1$ is growing. In the limiting case $\alpha\to 0$ the connection weight $w(\sigma)$
becomes Dirac's delta function (\ref{kudeltafkt}).}
\end{figure}
The limiting value of (\ref{wagnera2g}) for $u\to \infty$ is determined with the help of
the expansion \cite[1.513]{grad}
\begin{equation}
\ln\frac{1+x}{1-x}=2\sum_{k=1}^{\infty}\frac{1}{2k-1}\,x^{2k-1}\,,\quad x^2<1\,,
\end{equation}
and reads $\displaystyle{\lim_{u\to\,\infty}\alpha=\gamma}$.
Thus, we conclude that the case $u\to\,\infty$  
corresponds to the instability point $\alpha_c=f_1^T f_1^R$, 
which was obtained from the linear stability analysis in 
Sec.~\ref{linanalys}. Correspondingly, using again (\ref{wagnera2g}), we observe
that the connection weight (\ref{117}) coincides in the limit $u\to \infty$ with
a uniform distribution: 
$\displaystyle{\lim_{\alpha\uparrow \alpha_c} w(\sigma)=1}\,.$
Another biological important special case is $u\downarrow 1$, where we obtain
from the transcendental relation (\ref{wagnera2g}) 
$\displaystyle{\lim_{u\downarrow 1}\alpha=0}\,.$
Furthermore, considering the limit $u\downarrow 1$ in (\ref{117}) for $\sigma\not=u$, we obtain $w(\sigma)\to 0$.
On the other hand, integrating (\ref{117}) for $u\downarrow 1$ over $\sigma$ yields 2.
Therefore, we conclude that the connection weight $(\ref{117})$ becomes 
in this limit Dirac's delta function:
\begin{equation} \label{kudeltafkt}
\lim_{\alpha\downarrow 0} w(\sigma)=4\delta(\sigma-1)\,.
\end{equation}
Thus, decreasing the control parameter $\alpha$ means that the projection
between two spheres becomes sharper and sharper (see Fig.~\ref{alphavonu}b).
A perfect one-to-one retinotopy is achieved for $\alpha=0$ when the uniform and
undifferentiated formation of new synapses onto the tectum is completely terminated.

\section{Summary}

In this paper we have explicitly applied our generic model for the emergence of
retinotopic projections between manifolds of different geometry to
one- and two-dimensional Euclidean and spherical manifolds. By treating retina and
tectum as strings we generalized the original approach of H{\"a}ussler and
von der Malsburg where both were modelled as one-dimensional
discrete cell arrays. This change from discrete to continuous variables
is physiologically reasonable because of the high cell density in
vertebrate animals. By using a continuous instead of a discrete
model we emphasized that we are not interested in the dynamics of the
single cell but in the evolving global spatio-temporal patterns of the system.
Furthermore, continuous variables are helpful to describe retinotopic projections
between manifolds of different magnitudes as we have seen by the example of
two strings of different lengths. In case of discrete cell arrays with different
cell numbers it is not clear what a perfect retinotopy means, whereas in the
continuous case a perfect one-to-one projection can be described without
offending the bijectivity of the projections. Finally, as the one-dimensional
string model could only serve as a simplistic approximation to the real biological
situation, we have also investigated under which conditions retinotopic
projections between planar networks of neurons arise. Obviously, this increase
of the spatial dimension rendered the synergetic analysis so complicated that we
were only able to treat physiologically interesting special cases. While for
strings retinotopy was only possible for monotonically decreasing cooperativity
functions, we found that this is no longer true for planes.\\

Applying our generalized H{\"a}ussler-von der Malsburg equations \cite{gpw1} 
to strings and to spheres, led to remarkably analogous results.
Both for one-dimensional strings and for spheres we have
furnished proof that our generalized H{\"a}ussler-von der Malsburg equations describe,
indeed, the emergence of a perfect one-to-one retinotopy. Furthermore,
we have shown in both cases that the underlying order parameter
equations follow from a potential dynamics and do not contain
quadratic terms. However, in contrast to strings, spherical manifolds
represent a more adequate description for retina and tectum.
Therefore, the spherical case represents an essential progress in
the understanding of the ontogenetic development of neural connections
between retina and tectum.

\end{document}